\documentclass[10pt,sigplan,letterpaper,twocolumn]{article}
\usepackage[10pt,nocopyright]{sigmin}

\usepackage{listings}
\usepackage{color}
\usepackage[hyphens]{url}                                         
\usepackage[breaklinks,colorlinks]{hyperref}                      
\usepackage[usenames,dvipsnames]{xcolor}                          
\hypersetup{citecolor=blue,linkcolor=blue}                        
\usepackage{amsmath,amsopn,amssymb,amsthm}                        
\usepackage{thmtools}
\usepackage{thmbox}
\usepackage[toc,page]{appendix}
\usepackage{subcaption}                                           
\usepackage{endnotes,microtype,xspace,fancyvrb,multirow} 
\usepackage{epigraph} 

\usepackage{booktabs}      
\usepackage{tabularx}
\usepackage{array,underscore,relsize}                             
\usepackage[T1]{fontenc}                                          
\usepackage{times}                                                
\usepackage{fancyhdr,lastpage}                                    
\usepackage{enumitem}                                             
\usepackage[labelfont=bf,font=normal,skip=1pt]{caption}           
\usepackage[export]{adjustbox}                                    
\usepackage{breakurl}                                             
\usepackage{pifont}                                               
\usepackage{tablefootnote}                                        
\usepackage{bbding}                                        
\usepackage[linesnumbered,vlined,boxed,commentsnumbered]{algorithm2e}
\DontPrintSemicolon
\newcommand{\ra}[1]{\renewcommand{\arraystretch}{#1}}
\usepackage{amssymb}
\usepackage[normalem]{ulem}
\usepackage[hang,flushmargin]{footmisc}
\usepackage{textcomp}
\usepackage{graphicx}
\usepackage{authblk}

\usepackage[compact]{titlesec}
\titleformat*{\section}{\large\bfseries}
\titleformat*{\subsection}{\normalsize\bfseries}
\titleformat*{\subsubsection}{\normalsize\bfseries}
\titlespacing{\section}{0pt}{3ex}{1ex}
\titlespacing{\subsection}{0pt}{2ex}{1ex}

\hypersetup{
  colorlinks,
  linkcolor={red!50!black},
  citecolor={blue!50!black},
  urlcolor={blue!80!black}
}

\newcommand{\rkv}{{{R-KVS}}} 
\newcommand{\drtm}{{{DrTM-KV}}}

\newcommand{\dpu}{{{Bluefield}}}
\newcommand{\snic}{{{SmartNIC}}}
\newcommand{\etal}{\textit{et al}.}
\newcommand{\linefs}{LineFS}

\newcommand{\fig}[1]{Figure{~\ref{#1}}}

\newcommand{\one}{\texttt{\uppercase\expandafter{\romannumeral1}}}
\newcommand{\two}{\texttt{\uppercase\expandafter{\romannumeral2}}}

\def\ie{i.e.,~}
\def\eg{e.g.,~}

\newcommand{\refine}[1]{\textcolor{blue}{#1}}

\newcommand{\stitle}[1]{\vspace{1.1ex}\noindent{\bf #1}}

\usepackage{tikz}

\usepackage{balance}
\begin{document}

\title{\Large \bf Characterizing Off-path SmartNIC for Accelerating Distributed Systems}



\setlength{\affilsep}{0.5em}
\author[1,2]{Xingda Wei}
\author[1,2]{Rongxin Cheng}
\author[1]{Yuhan Yang}
\author[1,2]{Rong Chen}
\author[1] {Haibo Chen}
\affil[1]{\vspace{-2.mm}Institute of Parallel and Distributed Systems, SEIEE, Shanghai Jiao Tong University\vspace{0.8mm}}
\affil[2]{Shanghai AI Laboratory\vspace{-1.mm}}

\date{}
\maketitle

\frenchspacing

\begin{abstract}
\noindent
SmartNICs have recently emerged as an appealing device for accelerating distributed systems.
However, there has not been a comprehensive characterization of SmartNICs, and existing designs 
typically only leverage a single communication path for workload offloading. 
This paper presents the first holistic study of a representative off-path SmartNIC, 
specifically the Bluefield-2, from a communication-path perspective. 
Our experimental study systematically explores the key performance characteristics
of communication among the client, on-board SoC, and host, 
and offers insightful findings and advice for designers.
Moreover, we propose the concurrent use of multiple communication paths 
of a {\snic} and present 
a pioneering guideline to expose 
new optimization opportunities for various distributed systems. 
To demonstrate the effectiveness of our approach, 
we conducted case studies on a SmartNIC-based distributed file system (LineFS) 
and an RDMA-based disaggregated key-value store (DrTM-KV). Our experimental results 
show improvements of up to 30\% and 25\% for LineFS and DrTM-KV, respectively.
\end{abstract}

\section{Introduction}
\label{sec:intro}

\noindent
Remote Direct Memory Access (RDMA) has been widely adopted in modern data centers~\cite{DBLP:conf/sigcomm/GuoWDSYPL16,tsai2017lite,DBLP:conf/nsdi/GaoLTXZPLWLYFZL21}, 
pushing network bandwidth (towards 400\,Gbps~\cite{connect7}) and distributed system performance~\cite{farm,drtm-h,xstore,wukong,DBLP:conf/sosp/WeiSCCC15,xie2019pragh,krcore} 
to the next level. 
However, the high-speed network requires more CPU resources to saturate a fast RDMA-capable NIC (RNIC)~\cite{ipipe}, 
which places a significant CPU burden on distributed systems~\cite{linefs}. 
One-sided RDMA can alleviate CPU pressures by enabling the RNIC to directly read and write 
host memory in a CPU-bypass way.
However, the limited offloading capabilities may cause network amplifications 
and thus degrade system performance~\cite{xenic,DBLP:conf/usenix/KaliaKA16}. 

The continuous improvements in RDMA~\cite{cachecloud} and the essential power and memory walls 
of CPUs have led to the emergence of {\snic}s---the RNICs
with programmable capabilities.
These NICs offer systems the opportunity to offload 
more complex computations to the NIC.
Currently, there are two main types of {\snic}s. The first one is the \emph{on-path} {\snic}~\cite{liquidio},
which directly exposes the processing units (NIC cores) for handling RDMA packets to the systems. 
Unfortunately, programming low-level NIC cores with firmware~\cite{ipipe, xenic}
and isolating the offloaded program from normal RDMA requests pose significant burdens on developers. 
To simplify system development, 
the \emph{off-path} {\snic}~\cite{bluefield-2,bluefield-3,broadcom-stingray,innova}
attaches a programmable multicore SoC (with DRAM) next to the RNIC cores, 
which is off the critical path of RDMA.
Thanks to this separation, 
the SoC is independent of normal RDMA requests and can further deploy 
a full-fledged OS to make the developments easier~\cite{linefs}. 
Specifically, developers can treat the SoC as a separate server. 
In this paper, 
we focus on off-path {\snic}{\footnote{\footnotesize{This paper will use ``{\snic}'' 
(or ``SNIC'' for brevity) to specifically refer to off-path {\snic}s.}}} 
due to its generality and programmability.

There have been valuable studies on characterizing off-path {\snic}s~\cite{ipipe,
DBLP:journals/corr/abs-2105-06619,linefs,thostrupdbms,DBLP:conf/hotnets/AmaroLOKPRS20}, 
with a focus on their ability to offload computation.
A key finding is that the computing power of off-path {\snic}s is weaker than 
that of the host~\cite{ipipe, DBLP:journals/corr/abs-2105-06619, linefs}. 
This means that off-path {\snic}s do not improve the speed of a single network path, 
such as that between NIC and the host.  
For example, iPipe~\cite{ipipe} found that the path between the host and SoC has 
a relatively high latency due to the support for more developer-friendly RDMA. 

Although prior work has been valuable in utilizing {\snic}s for distributed systems, 
it has primarily focused on offloading computation to the {\snic}'s SoC. 
However, it is surprising that the fundamental function of {\snic}s, namely networking, 
has been overlooked despite its significant impact on overall performance.
In fact, networking on the {\snic} is intricate, because it provides 
multiple communication paths. 
For example, {\snic}s support using RDMA to access the memory of the host or SoC, 
as well as exchanging data between the host and the SoC.

To this end, this paper conducts the first systematic study on characterizing 
the performance of communication paths of {\snic}. 
Unlike previous studies that simply report basic performance numbers~\cite{DBLP:journals/corr/abs-2105-06619,linefs,thostrupdbms}, 
we systematically analyze the performance implications of {\snic} architecture 
on different paths. Specifically, we investigate 
why and when one path may be faster than another, identify the bottlenecks for each path, 
examine how the heterogeneity of the SoC brings performance anomalies in paths related to the SoC, 
and finally explore how paths interact with each other. 
The main highlights of our results are: 

\begin{itemize}[leftmargin=*,leftmargin=10pt,itemindent=0pt] 

    \item \emph{Different paths exhibit diverse performance characteristics.}   
    The RDMA path from the NIC to the SoC is up to 1.48$\times$ faster than 
	the path to the host. 

    \item \emph{The SoC introduces new performance anomalies to paths related to it}.  
    The low-level hardware details of the SoC, including the memory access path and 
	PCIe MTU, differ from those of the more powerful host CPU. 
    Without considering such factors, RDMA requests involving the SoC suffer from 
	up to 48\% bandwidth degradation. 
    
    \item \emph{The paths between the SoC and the host may underutilize the PCIe.}
    RDMA from the SoC to the host (and vice versa) crosses the NIC internal PCIe twice. 
	It can only utilize half of the PCIe bandwidth 
    and requires processing up to 6$\times$ more PCIe packets than the others.
    DMA only passes the PCIe once, 
    but it is not always faster than RDMA due to the weaker SoC DMA engine (compared to the one on the RNIC) and 
    also suffers from packet amplifications.
\end{itemize}

Based on our performance characterization, we found that prior approaches, 
which mainly optimize a single path for a specific functionality of distributed systems, 
failed to fully exploit {\snic}s. This is because a single path
cannot utilize the computing and networking capability of {\snic}s. 
Further, only considering a single path may ignore
resource interference between different paths (e.g., the PCIe and PCIe switches). 
As a result, {\linefs} can only utilize up to 117\,Gbps of bandwidth on a 200\,Gbps {\snic}. 
A similar issue exists in {\snic}-based disaggregated key-value store: while choosing 
a path to offload all key-value (KV) store operations to the {\snic} SoC 
can eliminate the network amplification in existing RDMA-based key-value stores, 
the wimpier computing power of {\snic} SoC limits its overall throughput. 

Based on the observations from our study, we further propose an optimization guideline to
help designers smartly exploiting multiple paths of {\snic}s. 
Instead of optimizing distributed systems along a single path, it 
holistically exploits multiple paths for functionalities with different characteristics 
and carefully considers 
cross-path interference. 
To demonstrate the efficacy of our guideline, 
we conduct two case studies by optimizing two state-of-the-art systems, 
namely {\linefs}~\cite{linefs} and DrTM-KV~\cite{drtm-tocs,drtm-h}. 
Due to the exposed new optimization spaces, following our guideline can
improve the performance of {\linefs} and DrTM-KV by up to 30\% and 25\% accordingly. 

\stitle{Contributions.} We summarize our contributions as follows:  
\begin{itemize}[leftmargin=*,leftmargin=10pt,itemindent=0pt] 

\item A comprehensive performance characterization of representative 
off-path {\snic}s, with a particular focus on various communication paths. 

\item The first optimization guideline 
for smartly exploiting the multiple paths of {\snic}s with
managed cross-path resource interference.  

\item Two case studies on SmartNIC-accelerated distributed systems (i.e., 
file system and key-value store) with notable performance improvements, 
demonstrating 
the efficacy of our guideline.  

\end{itemize}

\stitle{Assumptions and generalizability of our work.}
We assume an off-path {\snic} with the following architecture:
the SoC is linked with NIC cores via a PCIe switch, 
and there is heterogeneity between SoC and host CPUs.
We believe this is a representative architecture, as many older 
(e.g., NVIDIA Bluefield-1~\cite{bluefield-1}, Broadcom Stingray~\cite{broadcom-stingray}), 
current (e.g., NVIDIA Innova2~\cite{innova}, Bluefield-2~\cite{bluefield-2}), and 
upcoming {\snic}s (e.g., Bluefield-3~\cite{bluefield-3}, Marvell OCTEON 10 DPU~\cite{OCTEON}) 
use a similar setup.
We conducted experiments on 
Bluefield-2~\cite{bluefield-2}---the state-of-the-art 
{\snic} with this architecture.
Meanwhile, we also confirmed that our results hold for Bluefield-1.

However, 
we acknowledge that significant architectural changes (e.g., on-path {\snic}s) 
may affect our findings. Nevertheless, we argue that our methodology---first studying 
the performance implications of each communication path and then smartly exploiting 
multiple paths of {\snic}s---can be generalized to other {\snic}s.
Our benchmarking code, tools, and systems are available at 
{\small{\url{https://github.com/smartnickit-project}}}.

\begin{figure*}
\hspace{-10pt}
\begin{minipage}{1\linewidth}
\centering
\includegraphics[left, scale=1.1]{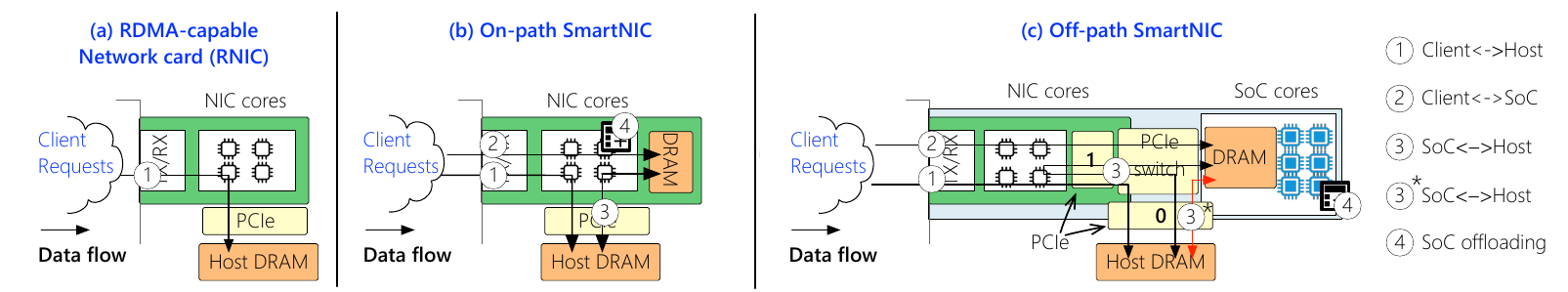}
\end{minipage} \\[10pt]  
\begin{minipage}{1\linewidth}
\caption{\small{%
Architecture of different NICs: 
(a) RDMA-capable NIC (RNIC), (b) on-path SmartNIC, and (c) off-path SmartNIC (our focus). 
}}
\label{fig:smartnic}
\end{minipage} \\[-20pt]  
\end{figure*}

\section{Background and Context}
\label{sec:bg}

\subsection{RDMA-capable NICs (RNICs)} 
\label{sec:rdma}

\noindent
RDMA (Remote Direct Memory Access) is a low-latency (2\,$\mu$s) and high-bandwidth (200\,Gbps)
network widely adopted in modern data centers~\cite{DBLP:conf/sigcomm/GuoWDSYPL16}.
One intuitive way to utilize RDMA is to accelerate message passing with its \emph{two-sided} 
primitives (SEND/RECV), such as {RDMA-based RPC~\cite{herd,farm,scalerpc,flock-rpc,erpc}}.
Alternatively, the one-sided primitives ({READ}/{WRITE}\footnote{\footnotesize{We use 
READ/WRITE to indicate RDMA READ/WRITE in this paper.}}) allow the RNIC
to access the host memory bypassing the host CPU. 
Specifically, the NIC core internally uses the direct memory access (DMA) feature 
of the PCIe link to access the host memory (see {\fig{fig:smartnic}(a)}).

Though RDMA has boosted the performance of 
many distributed systems~\cite{farm-v1,drtm-h,farm-v2,fasst}, usually by orders of magnitude,
it still has the following two problems 
especially when the RNICs scale up to higher performance.

\stitle{Issue\,\#1: Host CPU occupation.}
For two-sided primitives, distributed systems need 
non-trivial CPUs to saturate a powerful NIC.
Our measurements show that a 24-core server can only saturate
87 million packets per second (Mpps) on a 200\,Gbps RNIC (ConnectX-6), 
while NIC cores can process more than 195\,Mpps.%
\footnote{\footnotesize{Detailed hardware setups can be found 
in \textsection{\ref{sec:testbed}}.}}
A recent work further shows that a distributed file system 
requires 2.27$\times$ CPU cores to handle network packets,
when the network bandwidth scales from 25\,Gbps to 100\,Gbps~\cite{linefs}. 
Although deploying more powerful CPUs can alleviate this issue, 
RNIC bandwidth is also rapidly growing, 
currently reaching up to 400\,Gbps~\cite{connect7}.

\begin{figure}[!t]
\begin{minipage}{1\linewidth}
\centering\includegraphics[scale=.9]{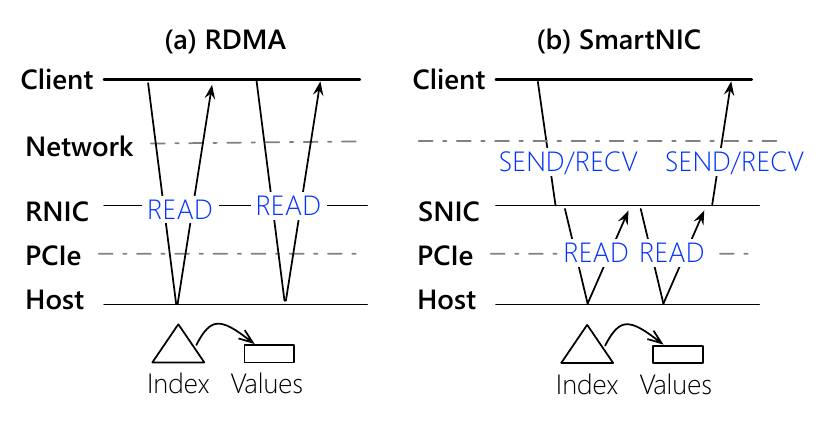}
\end{minipage} \\[5pt]
\begin{minipage}{1\linewidth}
\caption{\small{%
An illustration of a {get} request in a distributed in-memory key-value store
that is accelerated by using either (a) RNICs (w/ network amplification) 
or (b) SNICs (w/o network amplification).%
}}
\label{fig:bg-kv-offload-basic}
\end{minipage} \\[-10pt]
\end{figure}

\stitle{Issue\,\#2: Network amplification.}
Using one-sided RDMA primitives alleviates the host CPU pressure by 
allowing systems to offload memory accesses to the RNIC. 
However, the limited offloading capability constraints 
system performance, as a single request may involve multiple round trips 
of READs/WRITEs to complete (usually termed \emph{network amplification}).
{\fig{fig:bg-kv-offload-basic}}(a) exemplifies 
the execution of a \emph{get} request
on a distributed in-memory key-value store with one-sided RDMA READs. 
The client first uses one (or multiple) READ(s) to query the index for a given key. 
Based on the index returned by the previous READs, 
an additional READ is issued to retrieve the value. 

\subsection{From RNICs to SmartNICs}
\label{sec:snic}

\noindent
To address the limitations of RNICs, 
{\snic} adds an on-board memory (4--64\,GB)
together with various computation units (e.g., SoC) to the NIC.
By exposing them to the developers, 
{\snic} enables offloading customized computations onto it. 
Specifically, 
{\snic}s can be categorized as follows.

\stitle{On-path {\snic}.} 
As shown in {\fig{fig:smartnic}(b)}, the on-path {\snic} 
exposes the NIC cores to the systems with low-level programmable interfaces, 
allowing them to directly manipulate the raw packets. 
As the name implies, 
the offloaded code is \emph{on} the critical path of the network processing pipeline. 
Example NICs include Marvell LiquidIO~\cite{liquidio} and Netronome Agilio~\cite{agilio}.
The benefit is that the offloaded code is closer to the network packets.
Therefore, inline requests that only interact with the NIC, 
such as writing to the on-board memory (\ding{193}), 
are extremely efficient~\cite{ipipe,xenic}. 

However, on-path {\snic} has two limitations. 
First, the offloaded code (\ding{195}) competes NIC cores with the network requests sent 
to the host (\ding{192}). 
If offloading too much computation onto it, 
the normal networking requests sent to the host would suffer a significant degradation~\cite{ipipe}. 
Second, programming on-path NICs is difficult due to its low-level interface. 

\stitle{Off-path {\snic}.} 
As shown in {\fig{fig:smartnic}(c)}, the off-path {\snic} 
offers an alternative: 
it packages additional compute cores and memory in a separate SoC next to the NIC cores.
Therefore, 
the offloaded code is \emph{off} the critical path of the network processing pipeline. 
From the NIC perspective, the SoC can be viewed as a second full-fledged host
with an exclusive network interface. 
To bridge the NIC cores, SoC and host together, 
a PCIe switch is integrated inside the SmartNIC to properly dispatch network packets. 
Example NICs include NVIDIA Bluefield~\cite{bluefield-2,bluefield-3} and 
Broadcom Stingray~\cite{broadcom-stingray}. 

Compared to the on-path counterparts, 
the offloaded code does not affect the network performance of the host
as long as it does not involve network communications (\ding{193}). 
Thanks to this clear separation, 
the SoC can run a full-fledged kernel (\eg{Linux}) with a full network stack (\ie{RDMA}),
simplifying system development and allowing for offloading complex 
tasks~\cite{linefs}.
However, accelerating distributed systems with off-path {\snic}s 
is typically more challenging than using the on-path counterparts.
This is because the PCIe switch prolongs all communication paths 
(\ie{\ding{192}, \ding{193}, and \ding{194}}), 
causing potential performance degradation. 

\subsection{Target {\snic}: NVIDIA Bluefield-2}
\label{sec:bluefield}

\noindent
We conduct our study on Bluefield-2, 
a typical off-path {\snic} optimized for offloading general-purpose computations.
{\fig{fig:smartnic}}(c) illustrates its overall hardware architecture,
with detailed hardware configuration shown in Table~\ref{tab:bluefield-info}.

\begin{table}[!t]  
\vspace{3mm}
\centering    
\begin{minipage}{1\linewidth}
\caption{{\small{%
Hardware description of {\dpu}-2~\cite{bluefield-2}.
}}}        
\label{tab:bluefield-info}
\end{minipage} 
\begin{minipage}{1\linewidth}
\ra{1.1}
\small{            
\resizebox{1.\textwidth}{!}{%
\begin{tabular}{@{~}l@{~~~~}l@{~}}
\toprule
\textbf{Component}  & \textbf{Hardware description}                           \\ 
\midrule
NIC cores           & ConnectX-6 (2$\times$~100\,Gbps RDMA ports)    \\
SoC cores           & ARM Cortex-A72 processor (8\,cores, 2.75\,GHz)  \\
SoC memory          & 1$\times$~16\,GB of DDR4-1600 DRAM             \\
PCIe1               & PCIe 4.0~$\times$16 (256\,Gbps bandwidth)      \\
\bottomrule
\end{tabular}                    
}
} 
\end{minipage} \\[-10pt]  
\end{table}   

\begin{table*}[!t]  
\vspace{2.2mm}
\centering    
\begin{minipage}{1\linewidth}
\caption{{\small{%
Machine configurations in our two rack-scale RDMA-capable clusters. 
}}}        
\label{tab:hardware-info}
\end{minipage} 
\begin{minipage}{1\linewidth}
\ra{1.1}
\centering 
\small{            
\resizebox{1.\textwidth}{!}{%
\begin{tabular}{@{~}l@{~~~~}l@{~~~~}l l l l@{~}}
\toprule
\textbf{Name} & \textbf{Nodes} & \textbf{RDMA-capable NIC} 
& \textbf{Host PCIe (PCIe0)} & \textbf{Host CPU}  & \textbf{Host Memory}        \\ 
\midrule
SRV & 3              
&\begin{tabular}[c]{@{}l@{}} 1$\times$~ConnectX-6 ({200\,Gbps}) \\ 1$\times$~Bluefield-2 ({200\,Gbps})\end{tabular} 
& PCIe 4.0 $\times$16  ({256\,Gbps})     
& 2$\times$~Gold 5317 v4 (12\,cores, 3.6\,GHz) 
& 128\,GB DDR4-2933 \\
\midrule
CLI & 20             
& 1$\times$~ConnectX-4 (100\,Gbps)
& PCIe 3.0 $\times$16 (128\,Gbps)      
& 2$\times$~E5-2650 v4 (12\,cores, 2.2\,GHz) 
& 96\,GB DDR4-1600 \\
\bottomrule
\end{tabular}          
}} 
\end{minipage} \\[-5pt]  
\end{table*}  

\stitle{Hardware.}
{\dpu}-2 equips a mature RNIC (ConnectX-6) as its NIC cores for high-speed networking.
These cores support all RDMA operations.
Its programmability comes from an integrated on-borad SoC, 
which has 16\,GB DRAM and an ARM Cortex-A72 (8\,cores, 2.75\,GHz).
A PCIe 4.0 switch bridges the NIC cores, SoC and host together, 
enabling bi-direction data transfer of up to 256\,Gbps. Note that the SoC is linked to the PCIe switch 
via an internal link, rather than through PCIe.\footnote{\footnotesize{This 
has been confirmed 
by the NIC vendor.}}
Specifically,
the hardware counters provided by {\dpu}~\cite{bluefield-counter} also imply that 
it has only two 
PCIe 
links:
one linking RNIC with the switch (PCIe1) 
and the other linking the switch with the host (PCIe0). 

\stitle{Software.}
The SoC runs a full-fledged Linux, allowing developers to treat it as a normal ARM server. 
The kernel also hosts a full RDMA stack, making it convenient for enabling RDMA-based 
communication.
In addition, {\dpu} provides DOCA~\cite{doca} SDK for advanced usage, such as DMA.

\stitle{Communication primitives: RDMA and DMA.}
All communication paths related to the SoC are conducted using RDMA to simplify 
system development. As shown in {\fig{fig:smartnic}}(c),
clients can issue one-sided or two-sided RDMA requests to the SoC (\ding{193}),
similar to a twin server on the host. 
Meanwhile, the SoC can also interact with the host via RDMA, and vice versa (\ding{194}). 
However, exchanging data between the SoC and the host 
must pass through the RNIC (PCIe1 and NIC cores) {for RDMA support}, 
which adds a hidden bottleneck to this path.
Fortunately, we found that {\dpu} further provides DMA support (\ding{194}$^*$) 
with DOCA~\cite{doca}, allowing the SoC to use DMA to access the host memory (and vice versa),
bypassing the RNIC.

\stitle{Existing state of exploring {\dpu}.}
Previous studies~\cite{ipipe,DBLP:journals/corr/abs-2105-06619,linefs,thostrupdbms} 
have mainly focused on the computing power of {\dpu} (\ding{195} in {\fig{fig:smartnic}}),
revealing the relative weakness of the SoC cores in terms of performing offloaded tasks 
and sending network requests. This is because the frequency and number of cores 
are inferior to those of the host CPU.
Due the power constraints of {\snic}s,
it is unlikely that the relative performance comparison 
between the NIC and host CPU will change.
Hence, we take this as a premise during our investigation.

In contrast, few studies have considered various communication patterns in {\dpu} 
(i.e., \ding{192}, \ding{193}, and \ding{194}), which are the main focus of our work.
Thostrup~{\etal}~\cite{thostrupdbms} found that 
accessing the SoC memory (\ding{193}) using READ is faster than 
accessing the host memory (\ding{192}) in the same way. 
iPipe~\cite{ipipe} shows that using RDMA to communicate between the host 
and SoC (\ding{194}) has high latency due to the software overhead of supporting RDMA.
This paper systematically explores the performance characteristics 
of {\dpu} and summarizes insightful lessons and advice for future system developers.

\subsection{Notation and testbed}
\label{sec:setup}

\vspace{-2mm}
\stitle{Notations.}
This paper follows {\dpu}'s hardware specification when describing low-level hardware details 
related to {\dpu}-2. As shown in {\fig{fig:smartnic}}(c),
``PCIe1'' refers to the PCIe link connecting the NIC cores to the PCIe switch, 
and ``PCIe0'' refers to the link connecting the switch and the host's PCIe controller. 
The ARM cores, along with the on-chip memory of {\dpu}-2, are collectively referred to as ``SoC.''
The machine hosting Bluefield-2 is referred as the ``host.''
Furthermore, we use the terms ``requester'' and ``responder'' to refer to 
the machine issuing the RDMA requests and the destination hardware component, respectively.
For example, in \fig{fig:smartnic}(c), the requesters of paths \ding{192} and \ding{193} 
are any RDMA-capable machines (also called clients), and the responders are the {host} 
and {SoC}, respectively. 
For path \ding{194}, the requester and responder are the host and SoC, respectively, 
and vice versa. 

\label{sec:testbed}
\stitle{Testbed.}
Table~\ref{tab:hardware-info} presents the machine configurations in our testbed.
To best utilize {\snic}, we deploy {\dpu}-2 on the servers (SRV) with matching 
PCIe link (PCIe 4.0) by default. 
These machines can replace {\dpu}-2 with 200\,Gbps ConnectX-6 (RNIC) for comparisons. 
Other machines (CLIs) serve as clients that issue RDMA requests to the servers. 
All machines in SRV and CLIs are connected through a Mellanox SB7890 100\,Gbps 
InfiniBand Switch. 
Note that the network performance of the evaluated 200\,Gbps NIC is not limited 
since they connect to the switch with two 100\,Gbps ports.

\begin{table}[th]  
    \vspace{3.mm}
    \centering    
    \begin{minipage}{1\linewidth}
    \caption{{\small{%
	The findings and advice from our study. 
	Claims supported by sufficient evidence are denoted by \textbf{E}, while 
	those supported by hypotheses are denoted by \textbf{H}.%
    }}}        
    \label{tab:summary}
    \end{minipage} \\[2pt]  
    \begin{minipage}{1\linewidth}
    \ra{1.1}
	\centering
    \small{            
    \resizebox{1.\textwidth}{!}{%
    \begin{tabular}{@{~}l@{~~~}l@{~~}c@{~}}
        \toprule
        \textbf{SNIC Paths} & \textbf{Findings/Advice} & \textbf{E/H} \\ 
		\midrule
        \ding{192} (\textsection{\ref{sec:dataflow-1}})           
		& Throughput of RDMA is lower than RNIC                & H   \\ 
		& Latency of RDMA is higher than RNIC                  & E   \\ 
		\midrule
        \ding{193} (\textsection{\ref{sec:dataflow-2}})           
		& One-sided RDMA performance is better                 & H   \\
        & Avoid skewed memory accesses                         & E   \\
        & Avoid large READ requests                            & H   \\ 
		\midrule
        \ding{194}/\ding{194}* (\textsection{\ref{sec:dataflow-3}})           
        & RDMA overuses the PCIe bandwidth                     & E   \\
        & Avoid large READ/WRTIE requests                      & H   \\
        & Enable doorbell batching carefully for RDMA          & E   \\ 
		& Use DMA (\ding{194}*) to improve PCIe utilization    & E   \\
		\midrule
        \ding{192}+\ding{193} (\textsection{\ref{sec:dpu-concurrent}})           
		& Improve throughput by using paths \ding{192} and \ding{193}   & H \\
		& concurrently (esp. in opposite directions)                    &   \\ 
        \ding{192}/\ding{193}+\ding{194} (\textsection{\ref{sec:dpu-concurrent}}) 
		& Selectively offload traffic to \ding{194}            & E \\
        \bottomrule
        \end{tabular}
    
    }} 
    \end{minipage} \\[-8pt]  
    \end{table}  

\begin{figure*}[!t]
  \hspace{-2mm}
  \centering 
  \includegraphics[scale=1.2]{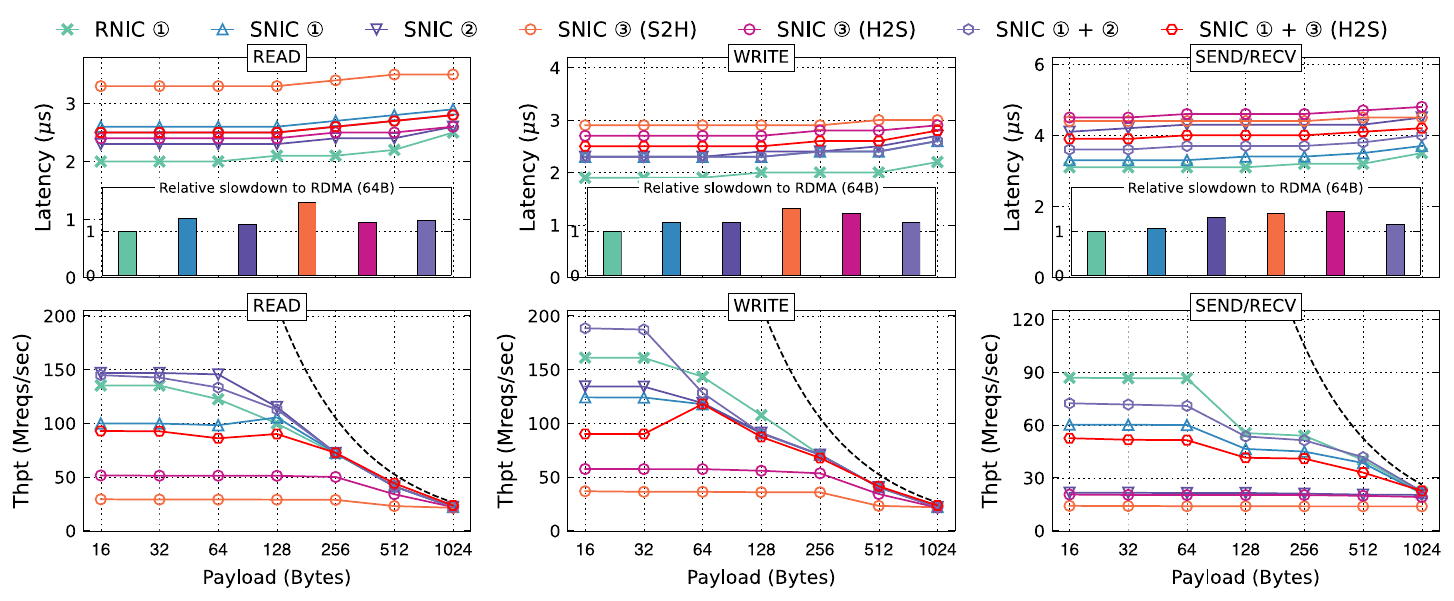}  \\[5pt]
  \begin{minipage}{1\linewidth}
  \caption{\small{%
  The end-to-end latency (upper) and peak throughput (lower) of random inbound RDMA requests 
  on different NICs. The symbols \ding{192}, \ding{193}, and \ding{194} in the legend 
  correspond to the communication paths listed in {\fig{fig:smartnic}}.
  }}
  \label{fig:study-thpt-lat}
  \end{minipage} \\[-5pt]
  \end{figure*} 

\section{Characterizing SmartNIC Performance}
\label{sec:dpu-study}

\noindent
As mentioned in \textsection{\ref{sec:bluefield}}, 
it is well-known that the \emph{computing power} of NIC is wimpier than 
that of the host CPU. 
Therefore, we focus on analyzing the \emph{communication efficiency} of {\snic}.
{\fig{fig:study-thpt-lat}} shows the end-to-end latency and peak throughput of sending  
different RDMA requests (e.g., READ, WRTIE, and SEND/RECV) using either RNIC or {\snic} 
through different communication paths.

\stitle{Evaluation setup.}
We conducted our experiments on the clusters described in Table~\ref{tab:hardware-info}, 
using a state-of-the-art RDMA communication framework~\cite{drtm-h}. 
For one-sided operations (READ and WRITE), the requester communicates with one responder 
using RDMA's reliable connection (RC) queue pairs (QPs).
The responder addresses are randomly chosen from a 10\,GB address space by default. 
For two-sided operations (SEND/RECV), 
the responder implements an echo server that utilizes all available cores 
for handling messages, and the requester communicates with it 
via unreliable datagram (UD) QPs for better performance~\cite{fasst,drtm-h,erpc}. 
For end-to-end latency, 
we deploy one requester machine to prevent interferences from queuing effects.
For peak throughput, 
we use up to eleven requester machines to saturate the responder.
Finally, we enable all well-known optimizations, including address alignment~\cite{DBLP:conf/usenix/WeiX00Z21}, 
unsignaled requests~\cite{herd} and huge pages~\cite{farm} to
prevent side effects from misusing RDMA.

\subsection{Communication from Client to Host (path {{\ding{192}}})}
\label{sec:dataflow-1}


\begin{figure}[!t]
\vspace{-2mm}
\begin{minipage}{1\linewidth}
\hspace{-3mm}
\centering    
\includegraphics[scale=0.93]{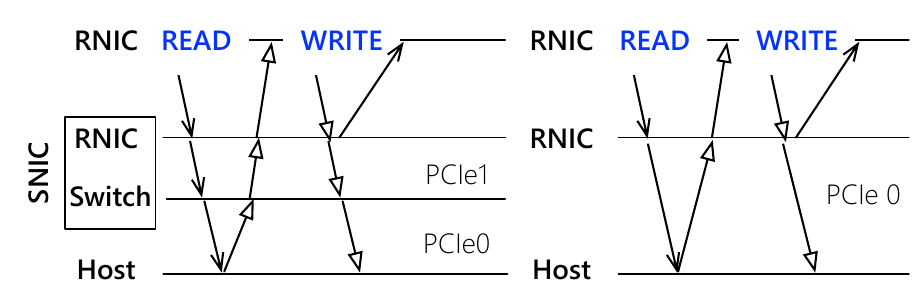}
\end{minipage} \\[8pt]
\begin{minipage}{1\linewidth}
\caption{\small{%
The exec. flow of READ/WRITE on SNIC and RNIC.
}}
\label{fig:dpu-slowdown}
\end{minipage} \\[-20pt]
\end{figure}

\vspace{-2mm}
\stitle{Latency.}
To compare communication with the host, 
we conduct an apple-to-apple comparison between {\dpu}-2 (SNIC \ding{192}) with 
ConnectX-6 (RNIC \ding{192}), as they share the same NIC cores~\cite{bluefield-2}.
Their performance gap best illustrates the ``performance tax'' paid by 
the {\snic} architecture. As shown in {\fig{fig:study-thpt-lat}},
SNIC \ding{192} has 15--30\%, 15--21\%, and 6--9\% higher latency than RNIC \ding{192}
for READ, WRITE, and SEND/RECV, respectively. 
The increased latency on SNIC comes mainly from the PCIe switch 
and PCIe1 between the host and NIC cores. 
The one-way PCIe latency is approximately 300\,ns, which is non-trivial for 
small RDMA requests (1--2\,$\mu$s). Note that the result is measured indirectly. 
Specifically, the end-to-end read latency on SNIC and RNIC is 2.6\,$\mu$s and 2.0\,$\mu$s, 
respectively.
Compared to RNIC, READ on SNIC passes through the PCIe switch
twice (see {\fig{fig:dpu-slowdown}}). 
Thus, the cost of each pass is around 300\,ns, which matches the number 
reported in recent literature~\cite{pcie-switch-lat}.
Furthermore, the increased latency of WRITE on SNIC is lower than that of READ, 
because it omits one pass through PCIe switch for completion~\cite{DBLP:conf/sigcomm/NeugebauerAZAL018}.
The latency of SEND/RECV on SNIC also increases, but mainly due to the larger CPU costs 
at the responder; the latency to post a request (via MMIO) on SNIC is higher than RNIC 
(399\,cycles vs. 279\,cycles).

\begin{figure*}[!t]
\hspace{-7pt}
\begin{minipage}{.25\textwidth}  
\vspace{-5pt}
\includegraphics[scale=1.2]{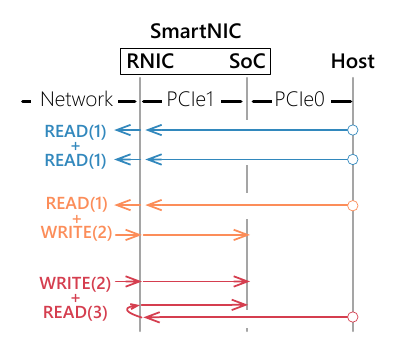}
\end{minipage}%
\begin{minipage}{.73\textwidth}
\hspace{10pt}
\includegraphics[scale=.95]{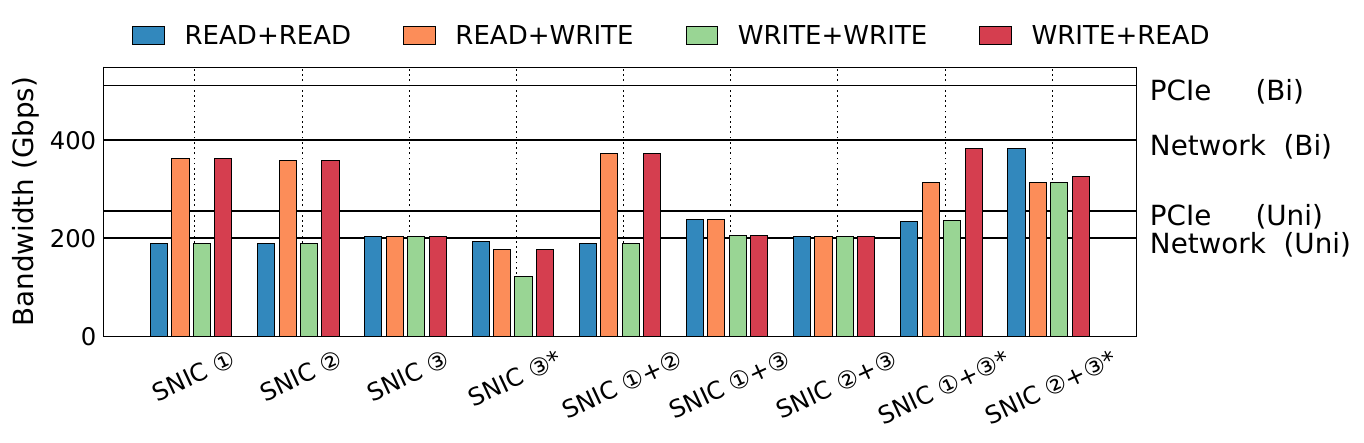}
\end{minipage}  \\[5pt]
\begin{minipage}{1\textwidth}
\caption{\small{%
(a) An illustration of three examples of different combinations 
of data flows on different paths,
and (b) the peak throughput of different combinations of data flows 
on different communications paths.
}}
\label{fig:offloading-bottlenecks}
\end{minipage} \\[-15pt]
\end{figure*}

\stitle{Throughput.}
As shown in {\fig{fig:study-thpt-lat}}, 
for READ, WRITE, and SEND/RECV, SNIC \ding{192} has 19--26\%,
15--22\%, and 3--36\% lower throughput than RNIC \ding{192}
for payloads less than 512\,bytes, respectively.  
We suspect the lower throughput is due to the longer latency in processing 
RDMA requests caused by PCIe switch.
However, for larger requests, the results are similar to using RNIC 
as both are bottlenecked by the network bandwidth.

\stitle{Bottleneck.}
The lowest bandwidth limit of NIC, PCIe1, and PCIe0 will first become the bottleneck 
for communication from client to host. 
On our testbed, the bottleneck is the network: 200\,Gbps.
On the other hand,
we find an interesting phenomenon: the total inbound bandwidth of the
requester can approach \emph{twice} the limit---400\,Gbps---because
the links are \emph{bi-directional}~\cite{rdma-bidirectional}.
Specifically, if packets flow in opposite directions, 
\eg{the READ and WRITE packets in {\fig{fig:offloading-bottlenecks}}(a)}, 
they can be multiplexed on the same link. 
To illustrate this, we dedicate two requesters (each with 12 threads to saturate the one-way bandwidth) 
to issue 4\,KB packets.
As shown in {\fig{fig:offloading-bottlenecks}}(b), 
if two clients send READ and WRITE requests separately, 
a total of 364\,Gbps bandwidth is measured on a 200\,Gbps NIC (see READ+WRITE of SNIC \ding{192}).
In contrast, if both clients send the same type of requests 
(either READ or WRITE), only about 190\,Gbps is measured. 
Note that though this phenomenon is widely known in traditional networking (i.e., messaging), 
where the messages are typically two-sided, 
it is largely ignored by many RDMA-based systems, because RDMA request can be one-sided.
 

\stitle{Takeaways.} 
Being ``smart'' incurs performance degradation for communicating with 
the host for small requests. 
For small requests,
we demonstrate that extending RNIC (ConnectX-6) to SNIC ({\dpu}-2) 
causes performance degradation by up to 36\% and 30\% in throughput and latency, 
respectively.
In general, for distributed systems that only use the path \ding{192}, it is recommended 
to use RNIC.
Although the overhead may be negligible for large requests or for networking with 
longer latency, RNIC is cheaper and more energy-efficient than SNIC.

\subsection{Communication from Client to SoC (path {{\ding{193}}})}
\label{sec:dataflow-2}

\vspace{-2mm}
\stitle{Latency.}
For sending requests from the client to SoC (SNIC \ding{193} in {\fig{fig:study-thpt-lat}}),
the latency of READ decreases by up to 14\% compared to the host (SNIC \ding{192}).
The \emph{reason} is that it skips PCIe0.
Yet, it is still 4--15\% higher than RNIC,
because requests still must go through the PCIe switch at PCIe1. 
For WRITE, SNIC \ding{193} provides similar performance as SNIC \ding{192} due to 
the asynchronous completion of cores (see \fig{fig:dpu-slowdown}).
For SEND/RECV, SNIC \ding{193} has 21--30\% higher latency than SNIC \ding{192}
due to the weaker computing power of SoC.

\stitle{Throughput.}
SNIC \ding{193} has better throughput than SNIC \ding{192}, reaching 1.08--1.48$\times$ 
for payloads less than 512\,bytes. 
Interestingly, the READ of SNIC \ding{193} is even 
higher than that of RNIC \ding{192} before reaching the peak network bandwidth.
For this undocumented results, we suspect that it is due to the closer packaging
of SoC memory and the PCIe switch.
Specifically, the SoC is linked to the PCIe switch via an internal link, 
rather than through PCIe.
Note that a confident analysis relies on the hardware details of {\dpu}, which 
unfortunately are not available now. 
For WRITE, SNIC \ding{193} is still lower than that of the RNIC \ding{192}.
Our \emph{hypotheses} are twofolds. 
First, SoC has fewer DRAM channels compared to the host (1 vs. 4), limiting 
the concurrency of write accesses. 
Nevertheless, READ is not affected because read accesses on DRAM are faster than write accesses~\cite{DBLP:conf/hpca/HassanVKGCPLEM17, DBLP:conf/micro/WangKJ22}. 
Second, SoC can only utilize a portion of NIC cores (see \textsection{\ref{sec:dpu-concurrent}}).
Finally, SEND/RECV has a poor performance on \ding{193}: 
it just achieves up to 64\% of the host (SNIC \ding{192}). 
This is due to the wimpy computing power of SoC, since the throughput of SEND/RECV 
is bottlenecked by the responder CPU to send the reply.

\stitle{Bottleneck.}
As shown in {\fig{fig:smartnic}}(c),
since SNIC \ding{193} only flows through NIC and PCIe1,
the bottleneck is their lower bandwidth limit, 
which is still {\dpu}-2's 200\,Gbps NIC. 
Therefore, as shown in {\fig{fig:offloading-bottlenecks}}(b), 
the performance of SNIC \ding{193} is the same as that of SNIC \ding{192},
namely the total of 400\,Gbps and 200\,Gbps bandwidth 
for opposite direction and same direction communication, respectively.

\begin{figure}[!t]
\vspace{-2mm}
\begin{minipage}{1\linewidth}
\hspace{-3mm}
\centering    
\includegraphics[scale=1.04]{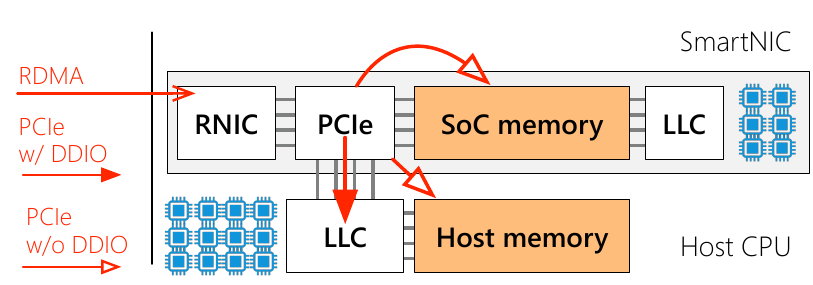}
\end{minipage} \\[10pt]
\begin{minipage}{1\linewidth}
\caption{\small{%
Different paths to access host and SoC memories.
}}
\label{fig:ddio}
\end{minipage} \\[-20pt]
\end{figure}

\vspace{5pt}
In addition to the basic RDMA performance of the SNIC, 
we found several factors that could also prevent distributed systems from 
achieving the aforementioned performance.

\stitle{Advice\,\#1: Avoid skewed memory accesses.}
The wimpy SoC cores may impact the memory access behavior of one-sided RDMA primitives,
because it usually supports fewer features compared to the more powerful host CPU cores. 
Specifically,
Data Direct I/O (DDIO)~\cite{ddio} is widely supported by the host CPUs, 
which allows the NIC to directly read/write data from/to its last level cache (LLC), 
as shown in {\fig{fig:ddio}}. 
SoC cores may also equip with similar features (e.g., ARM CCI~\cite{cci}), 
but whether to do so is vendor-specific. 
The SoC cores of our hardware
(ARM Cortex-A72 in {\dpu}-2) do not support DDIO.
We find that one-sided RDMA without DDIO is vulnerable to 
skewed memory accesses---the requested memory addresses fall into a small range.
This is \emph{because} DRAM requires a (not-too-small range) to utilize all memory modules concurrently.
LLC is faster than DRAM, so we suspect it can better tolerate the skewed access.

\begin{figure}[!t]
  \hspace{-2mm}
  \begin{minipage}{1\linewidth}
  \centering 
  \includegraphics[scale=1.1]{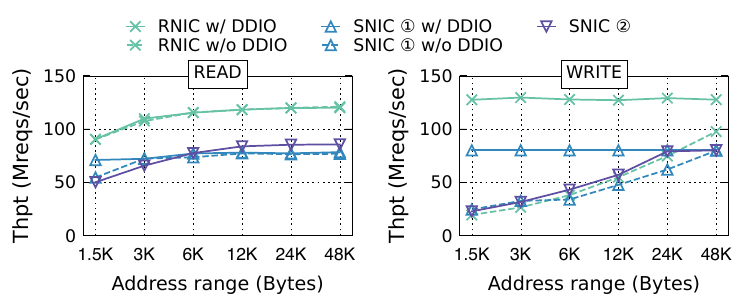}
  \end{minipage} \\[5pt]
  \begin{minipage}{1\linewidth}
  \caption{\small{%
  The peak throughput of accessing host memory and SoC memory via SNIC,
  READ (a) and WRITE (b).
  }}
  \label{fig:dca}
  \end{minipage} \\[-15pt]
  \end{figure} 

{\fig{fig:dca}} shows the peak throughput 
of accessing host memory and SoC memory via SNIC
with the increase of address ranges.\footnote{\footnotesize{Note that 
we attach {\dpu} to CLI machines for the evaluation because we are unable to 
disable DDIO on the SRV machines.}}
For WRITE, 
the throughput of SNIC \ding{193} using SoC 
drops to 22.7\,M reqs/s (from 77.9\,M reqs/s)
when address range decreases to 1.5\,KB (from 48\,KB).
In contrast, the performance of SNIC \ding{192} using Host CPU
is hardly affected when DDIO is enabled.
For READ, the degradation is relatively smaller. 
The throughput of SNIC \ding{193} drops from 85\,M reqs/s to 50\,M reqs/s 
when decreasing the range from 48\,KB to 1.5\,KB.
This is because 
DRAM can serve reads faster than writes~\cite{DBLP:conf/hpca/HassanVKGCPLEM17, DBLP:conf/micro/WangKJ22}. 
Finally, we also plot the RNIC results as a reference.
Under skewed workloads, we can see that \ding{192} also suffers a significant performance drop 
on WRITE when DDIO is disabled.

\begin{figure}[!t]
\hspace{-2mm}
\begin{minipage}{1\linewidth}
\includegraphics[scale=1.1]{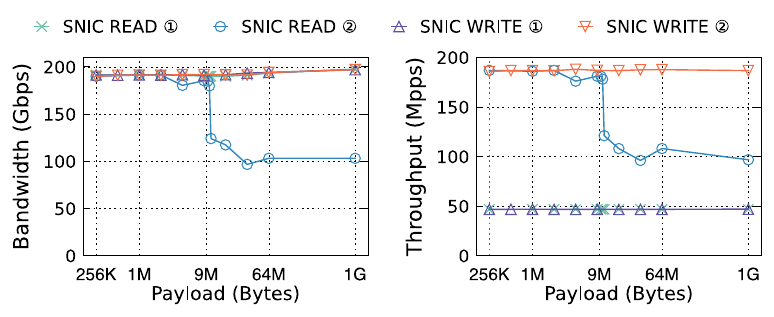}  
\end{minipage} \\[5pt]
\begin{minipage}{1\linewidth}
\caption{\small{%
The bandwidth (a) and PCIe packet throughput (b) for accessing (READ and WRITE)
the host (SNIC \ding{192}) and SoC (SNIC \ding{193}) via {\snic}. 
For brevity, we omit the result of SEND/RECV 
since it is the same as WRITE for large payloads~\cite{herd}.
}}
\label{fig:large-payload-analysis}
\end{minipage} \\[-10pt]
\end{figure}

\stitle{Advice\,\#2: Avoid large READ requests.}
It is common practice to use requests with large payloads to fully exploit network bandwidth. 
For example, using requests with payloads larger than 16\,KB is enough 
to saturate a 200\,Gbps RNIC even using a few threads.
Unfortunately, we observed that the READ performance of SNIC \ding{193} collapses
with request payload larger than 9\,MB, 
as shown in {\fig{fig:large-payload-analysis}}(a).
We \emph{suspect} that NIC cores suffer from head-of-line blocking 
when processing large READ requests. 
For a READ request, the NIC issues a PCIe read transaction to fetch the data, 
which is further segmented into multiple PCIe packets. 
The maximum size of a PCIe packet is determined by the PCIe Maximum Transfer Unit (MTU), 
negotiated by the linked hardware devices 
during bootstrap~\cite{DBLP:conf/sigcomm/NeugebauerAZAL018}.
Table~\ref{tab:pcie-payload} lists the PCIe MTU on our testbed.
SoC cores (the endpoint of SNIC \ding{193}) use a smaller PCIe MTU 
(128\,B) due to its weaker CPU.
As a result, 
NIC core that processes a large DMA read sent to SoC memory (SNIC \ding{193})
must wait for more PCIe packets to arrive, resulting in lengthy processing stalls. 
Since the overall NIC packet processing power is not the bottleneck: 
as shown in {\fig{fig:large-payload-analysis}}(b), 
the requests with payloads smaller than 9\,MB still can achieve a high processing rate 
while it collapses for the others, 
so we suspect some blocking happens at the NIC core.
Note that WRITE requests are not affected since DMA does not wait for 
the completion~\cite{pcie-talk,DBLP:conf/sigcomm/NeugebauerAZAL018}.

On the contrary, the host uses a larger PCIe MTU (512\,B), 
so it does not suffer from bandwidth degradation (SNIC {\ding{192}}). 
As shown in {\fig{fig:large-payload-analysis}}(b),
the NIC can issue 46.7 million PCIe packets per second 
to the host (SNIC {\ding{192}}).
The aggregated bandwidth reaches 191\,Gbps, bottlenecked by the network.

\begin{table}[!t]  
\vspace{3mm}
\begin{minipage}{1\linewidth}
\caption{{\small{
PCIe Maximum Transfer Unit (MTU) on our testbed, 
and the number of PCIe packets required to transfer $N$ bytes
via different communication paths of {\dpu}-2.
Our simplified model omits control-path packets 
(e.g., two-sided message arrival notification).	
}}}        
\label{tab:pcie-payload}
\end{minipage} \\[2pt]  
\begin{minipage}{1\linewidth}
\ra{1.1}
\centering
\small{            
\resizebox{.95\textwidth}{!}{%
\begin{tabular}{@{~}lc c c@{~}}
\toprule 
& & \textbf{Host CPU cores} ($H_{_{\text{MTU}}}$)  & \textbf{SoC cores} ($S_{_{\text{MTU}}}$) \\ 
\midrule
\textbf{PCIe MTU}  & & 512\,B  & 128\,B         \\
\bottomrule
\end{tabular}
}
} 
\end{minipage} \\[2pt]  
\begin{minipage}{1\linewidth}
\ra{1.1}
\centering
\small{            
\resizebox{.95\textwidth}{!}{%
\begin{tabular}{@{~}l ccc@{~}}
& \textbf{SNIC \ding{192}} 
& \textbf{SNIC \ding{193}} 
& \textbf{SNIC \ding{194}} \\ 
\midrule
\textbf{PCIe1}~~ 
&  $\lceil N/H_{_{\text{MTU}}}\rceil$    
&  $\lceil N/S_{_{\text{MTU}}}\rceil$       
&  $\lceil N/H_{_{\text{MTU}}}\rceil + \lceil N/S_{_{\text{MTU}}}\rceil $   \\
\textbf{PCIe0}~~ 
&  $\lceil N/H_{_{\text{MTU}}}\rceil$    
& --
&  $\lceil N/H_{_{\text{MTU}}}\rceil$                             \\
\bottomrule
\end{tabular}
}
} 
\end{minipage} \\[-10pt]  
\end{table}

\stitle{Takeaways.}
For READ and WRITE, sending requests to SoC is typically faster than 
that to the host (or even faster than via RNIC) because SoC is ``closer'' to the NIC 
(without PCIe0).
In contrast, using SEND/RECV 
to communicate with SoC is slower due to weak SoC cores. 
Furthermore,
designers still need to carefully consider the heterogeneity 
between host CPU cores and SoC cores to avoid performance anomalies.
Specifically, skewed memory accesses may degrade the performance 
due to the lack of DDIO support in SoC cores.
In addition, sending large READ requests to SoC may underutilize the bandwidth
so the request should be proactively segmented into smaller ones.

\subsection{Communication between SoC and Host (path {{\ding{194}}})}
\label{sec:dataflow-3}

\noindent
We first describe our measurements and findings of RDMA
and then compare RDMA (\ding{194}) with DMA (\ding{194}*). 

\stitle{Latency.}
As shown in {\fig{fig:study-thpt-lat}}, 
the latency of sending requests from SoC to the host (SNIC \ding{194} S2H) is very high, 
especially for READ, since the requester (SoC) takes longer to issue 
an RDMA request to the NIC. 
The latency in the opposite direction (from the host ot SoC, SNIC \ding{194} H2S) 
is reduced but still 4--17\% higher than SNIC \ding{193}. 
Although the intra-machine communication saves one network round-trip, 
it adds additional PCIe transfers. 
Specifically, the request on SNIC \ding{193} flows the requester-side PCIe (not shown 
in {\fig{fig:smartnic}}(c)), the network, PCIe1, and the PCIe switch, 
while the request on SNIC \ding{194} (H2S) flows PCIe0, the PCIe switch, 
PCIe1 twice (in and out), and the PCI switch (again).

\stitle{Throughput.}
For requests with payloads less than 512 bytes, the throughput of 
SNIC \ding{194} (both S2H and H2S) is dominated by the requester's capability to post networking requests. 
This is because a single requester machine (either SoC or the host) cannot saturate 
the NIC with small requests\footnote{\footnotesize{We use 
up to eleven requester machines for SNIC \ding{192} and SNIC \ding{193}.}}, 
the READ throughput of SNIC \ding{194} only reaches 29\,M reqs/s and 51.2\,M reqs/s
for S2H and H2S, respectively, still far from its limit.
For WRITE and SEND/RECV, the results are similar.
For larger requests, they are bottlenecked by the PCIe bandwidth,
which will be discussed in more detail next.

\stitle{Bottleneck.}
As shown in {\fig{fig:offloading-bottlenecks}}(b),
for packets flowing in a single direction, 
communication between host and SoC is bottlenecked by 
PCIe bandwidth (256\,Gbps) rather than the uninvolved NIC (200\,Gbps). 
Therefore, the peak bandwidth of SNIC \ding{194} is slightly higher than 
SNIC \ding{192} and \ding{193} (204\,Gbps vs. 191\,Gbps). 
Readers might be interested in why the results of SNIC \ding{194} 
cannot be close to 256\,Gbps. 
We \emph{suspect} that it requires much more PCIe packets than the others.
For packets flowing in opposite directions,
SNIC \ding{194} can not utilize
twice the limit as the other paths (i.e., SNIC \ding{192} and \ding{193}).
This is because RDMA overuses the PCIe: 
each request passes through PCIe1 twice (in and out),
exhausting the bi-directional link.

\begin{figure}[!t]
\begin{minipage}{1\linewidth}
\centering
\includegraphics[scale=1.1]{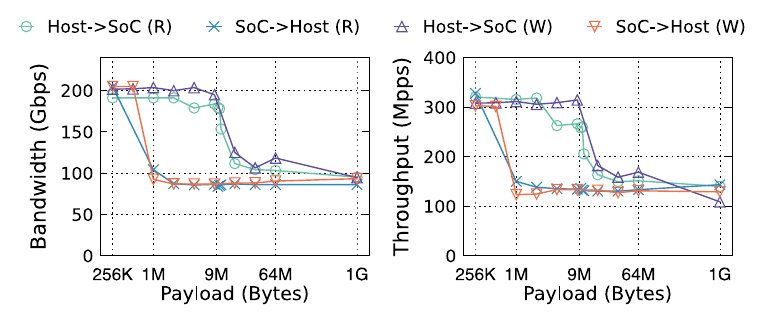}
\end{minipage} \\[5pt]
\begin{minipage}{1\linewidth}
\caption{\small{%
The bandwidth (a) and PCIe packets throughput (b) 
for sending (R)EAD/(W)RITE requests between the host and SoC. 
}}
\label{fig:large-payload-analysis-3}
\end{minipage} \\[-15pt]
\end{figure}

\stitle{Advice\,\#3: Avoid large READ/WRTIE requests.}
Communications between the host and SoC (SNIC \ding{194}) also suffers from 
bandwidth degradation for large READ requests like SNIC \ding{193},
possibly due to the head of line blocking as we have discussed before.
Moreover, 
this issue appears with large WRITE requests 
because the {\snic} must first read data from the requester and then write it
to the responder.
As shown in {\fig{fig:large-payload-analysis-3}}(a), the READ/WRITE performance of 
SNIC \ding{194} collapses to about 100\,Gbps for large requests.
Table~\ref{tab:pcie-payload} shows the number of PCIe packets required to 
transfer $N$ bytes via different communication paths. 
For SNIC \ding{194}, the NIC generates more packets due to passing through 
PCIe1 twice. Further, the performance of S2H collapses earlier than H2S 
as it will passes through PCIe1 first.
Suppose we transfer data at 200\,Gbps from SoC to the host.
The SoC cores first transfer 195\,M PCIe packets per second (pps) to the NIC (PCIe1), 
then the NIC forwards data back to the PCIe switch via PCIe1 again with 49\,Mpps 
(the host supports 512\,B MTU), 
and finally, the switch forwards 49\,Mpps through PCIe0.
Therefore, {\snic} should process at least 293\,Mpps for transferring 
data at 200\,Gbps, which is 3$\times$ and 1.5$\times$ higher than SNIC \ding{192} 
and SNIC \ding{193}, respectively. 
This is further confirmed by our measurements of the hardware counters. 
As shown in {\fig{fig:large-payload-analysis-3}}(b), 
for sending 256\,KB READ requests from SoC to the host, the bandwidth reaches 
204\,Gbps, and the NIC transfers about 320\,M PCIe packets per second. 


\begin{figure}[!t]
\begin{minipage}{1\linewidth}
\hspace{-10pt}
\centering 
\includegraphics[scale=1.1]{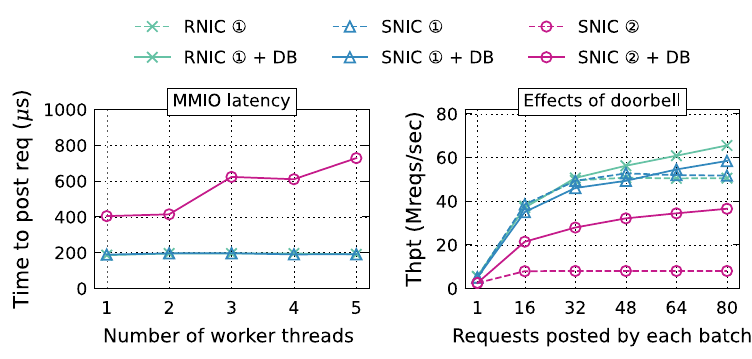}
\end{minipage} \\[5pt]
\begin{minipage}{1\linewidth}
\caption{\small{%
The latency of posting requests to NICs (a) and 
the impact of doorbell batching (DB) on the requester. 
}}
\label{fig:mmio-doorbell-soc}
\end{minipage} \\[-20pt]
\end{figure}

\stitle{Advice\,\#4: Enable doorbell batching carefully.}
The time of posting each request to the NIC is dominated by Memory-Mapped IO 
(MMIO)~\cite{drtm-h,DBLP:conf/usenix/KaliaKA16}. The SoC suffers a high MMIO latency when communicating 
with the host (see {\fig{fig:mmio-doorbell-soc}}(a)).
A known optimization is doorbell batching (DB)~\cite{DBLP:conf/usenix/KaliaKA16}:
to send a batch of $B$ requests, the requester first chains them together in memory, 
then use one MMIO to ask the NIC to read these requests with DMA in a CPU-bypass way.
DB reduces the number of MMIOs required from $B$ to $1$. 
Thus, for RNIC \ding{192} and SNIC \ding{193}, DB is always helpful and 
can bring 2--30\% performance improvement (see {\fig{fig:mmio-doorbell-soc}}(b)). 
For the communication between host and SoC (SNIC \ding{194}),
DB is still helpful at the SoC-side. As shown in {\fig{fig:mmio-doorbell-soc}}(b), 
when sending a batch of READs to the host, 
DB improves the SoC performance by 2.7--4.6$\times$ for batch sizes 16--80.
The huge improvement is partly due to the CPU-bypass feature of DMA,
and also because the NIC is faster in using DMA to read requests stored on SoC memory 
(see \textsection{\ref{sec:dataflow-2}}).
However, DB is not always helpful at the host-side, because it is slower to 
read host memory using NIC DMA (see \textsection{\ref{sec:dataflow-1}}). 
For batch sizes of 16, 32, and 48, DB decreases the throughput of host-SoC communication 
by 9\%, 7\%, and 6\%, respectively.

\begin{figure}[!t]
  \begin{minipage}{1\linewidth}
  \hspace{-1mm}
  \centering 
  \includegraphics[scale=1.1]{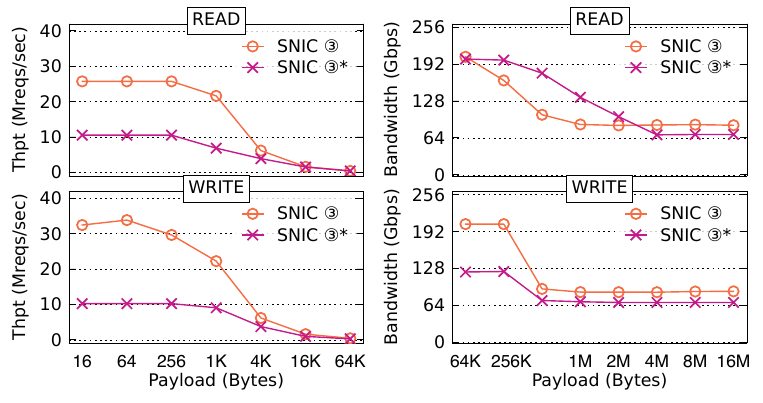}
  \end{minipage} \\[5pt]
  \begin{minipage}{1\linewidth}
  \caption{\small{%
  The throughput and bandwidth comparisons of using RDMA (\ding{194}) and DMA (\ding{194}*)
  when communicating from SoC to the host, the only supported primitive of DMA of our SNIC.
  }}
  \label{fig:rdma-vs-dma}
  \end{minipage} \\[-10pt]
  \end{figure} 

\stitle{RDMA (\ding{194}) vs. DMA (\ding{194}*).}
\label{sec:rdma-vs-dma}
Besides RDMA, SoC can use DMA (\ding{194}*) to read/write data from the host (and vice versa)
via the DMA engine inside the SoC.
It has the benefits of reducing 
two PCIe passes (PCIe1) and bypassing RNIC compared to RDMA (see {\fig{fig:smartnic}}),
resulting in a lower latency, e.g., 1.9\,$\mu$ vs. 2.6\,$\mu$ for 64\,B SoC to host READ.
However, we find the SoC DMA engine has a weaker processing power than RNIC (RDMA). 
For brevity, we only present results on SoC to host.
The results of host to SoC is the same as SoC to host since 
host DMAs are offloaded to SoC for execution~\cite{doca-dma-programming}.
As shown in \fig{fig:rdma-vs-dma}, for WRITE,
the peak throughput of DMA is only 47--59\% of that of RDMA for requests with payload less than 4\,KB.
The results of READ is similar.
DMA WRITE even fails to saturate the PCIe limit (256\,Gbps)
for payloads between 16\,KB and 1\,MB.
We suspect it is due to the poor processing capability of the SoC's DMA engine, 
yet we cannot confirm this without knowing the confidential internal design of the SoC.
Another observation from the bandwidth results is that 
DMA also suffers from the anomalies of RDMA (see Advice\,\#3):
For payloads larger than 1\,MB, 
there is a significant performance drop for both READ and WRITE. 

For bandwidth, \ding{194}* has a higher theoretical 
upper bound than \ding{194}:  
it is bottlenecked by the bidirectional bandwidth of PCIe, 
as it bypasses the PCIe1.
However, {\fig{fig:offloading-bottlenecks}} shows that it fails to achieve so (only 178\,Gbps for READ + WRITE).
This suggests that the slow DMA engine will first become the bottleneck.
Nevertheless, bypassing PCIe1 still has the benefits of reducing interferences to other paths.
We will discuss them in \textsection{\ref{sec:smart}} in detail.

\stitle{Takeaways.}
First, enabling doorbell batching is critical for SNIC \ding{194} at the SoC side,
because SoC has wimpy computation power. 
Yet, 
it is negatively impacted at the host side for small batch sizes.
Second, SNIC \ding{194} has a different bottleneck than SNIC \ding{192} and SNIC \ding{193}.
It is always bottlenecked by the uni-directional bandwidth of PCIe, 
while others are limited by the minimal bi-directional bandwidth of network and PCIe. 
If this factor is not adequately considered,
distributed systems will underutilize the NIC bandwidth
(see \textsection{\ref{sec:linefs}}).  
Third, though DMA utilizes PCIe better than RDMA for SoC to communicate with the host,
it has a lower throughput due to the weaker DMA engine at the SoC.
Finally, we should avoid transferring large requests between the host and SoC, for both RDMA and DMA 
and for both READ and WRITE.

\section{A Guideline for Smartly Exploiting Multiple Paths of {\snic}} 
\label{sec:smart}

\noindent
Previous approaches mainly leverage a single path of {\snic} to optimize 
a specific functionality of distributed systems. However, this cannot fully 
exploit the computing and networking capabilities of {\snic}s. 
Furthermore, only considering a single path may ignore interference on resources (e.g., PCIe and PCIe switch)
between different paths. Therefore, we first 
holistically study the performance characteristics of concurrently using multiple paths, 
and then lay out an optimization guideline for designers to smartly use {\snic}s. 

\subsection{Characterizing concurrent communication paths}
\label{sec:dpu-concurrent}

\vspace{-2mm}
\stitle{Concurrent communication with the host and the SoC (\ding{192}+\ding{193}).} 
We focus on the throughput results (see the lower part of {\fig{fig:study-thpt-lat}}) 
since the latency results are roughly the average of the two paths. 
We evaluate the peak throughput by assigning half of the clients to send requests 
to the host while the others to send to the SoC.
We can see that the total peak throughput of concurrently using \ding{192} and \ding{193}
(SNIC \ding{192}+\ding{193}) is typically faster than each of them. 
For READ, WRITE, and SEND/RECV, SNIC \ding{192}+\ding{193} outperforms
the lower of them by up to 1.45$\times$, 1.50$\times$, and 3.3$\times$, respectively.

\begin{figure}[!t]
\begin{minipage}{1\linewidth}
\centering 
\includegraphics[scale=1.05]{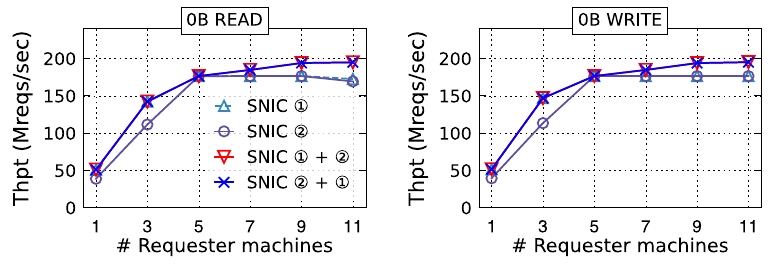}  \\[2pt]
\end{minipage} \\[5pt]
\begin{minipage}{1\linewidth}
\caption{\small{%
Throughput for (a) READ and (b) WRITE 
with the increases of requester machines.
}}
\label{fig:dpu-hybrid}
\end{minipage} \\[-15pt]
\end{figure}

For SEND/RECV, 
a concurrent path utilize both of the host and SoC to process the requests, 
so the performance improvement is clear.
However, the READ/WRITE performance improvement is non-intuitive and undocumented, 
since two paths should compete for NIC cores.
Our \emph{suspicion} is that 
the {\snic} internally reserves some NIC cores for each endpoint.
Therefore, sending requests to the host and the SoC concurrently can further increase 
\emph{peak} throughput by enabling more NIC cores.
To quantify this, 
we design a microbenchmark that 
first increases the requester machines to saturate the NIC and then 
changes the responder, as shown in {\fig{fig:dpu-hybrid}}.
All requests use 0\,B payload to avoid interference of DMA,
i.e., the request will return before passing PCIe1~\cite{ib}.
For READ, five requester machines are sufficient to saturate NIC cores 
when using SNIC \ding{192} or SNIC \ding{193} alone. 
Therefore, for concurrently using SNIC \ding{192} and SNIC \ding{193},
we first dedicate five requester machines for one responder, 
and then add requesters for the other responder.
Both cases (SNIC \ding{192}+\ding{193} and SNIC \ding{193}+\ding{192}) 
offer similar performance, with 4--13\% and 5--10\% higher throughput 
than using SNIC \ding{192} or SNIC \ding{193} alone.
For WRTIE, all results are almost the same.

Finally, as expected, the aggregated throughput of the two paths (SNIC \ding{192} 
and SNIC \ding{193}) is much higher than concurrently using them (352\,Mpps vs. 
195\,Mpps), indicating that most NIC cores are still shared, i.e., each can 
communicate with two endpoints, and only a few is dedicated. 
This also implies that concurrently using multiple resources of {\snic} is non-trivial.

\stitle{Concurrent inter- and intra-machine communication 
(\ding{192}/\ding{193}+\ding{194}).}
There exist four concurrent combinations of inter- and intra-machine communication.
For brevity, 
we focus on the results of SNIC \ding{192}+\ding{194}/H2S,
other combinations are similar.
To study the concurrent usage of the two paths, 
we first deploy sufficient clients (five requester machines) to saturate the network 
for SNIC \ding{192}.
Afterward, we start the requester on the host (one machine with 24 threads)
sending RDMA requests to the SoC (SNIC \ding{194}/H2S).
Our measurements reveal that concurrently enabling intra-machine communication degrades 
the performance of inter-machine communication. 
As shown in {\fig{fig:study-thpt-lat}}, 
for READ, WRITE, and SEND/RECV, the throughput of small requests (less than 512 bytes) 
drops 7--15\%, 4--27\%, and 9--14\%, 
by comparing SNIC \ding{192} and SNIC \ding{192}+\ding{194}(H2S). 
For large requests, the performance is always bottlenecked by the network bandwidth, 
so the degradation is negligible.

The SNIC \ding{194} affects other communication paths 
of {\snic}, because it relies on the NIC (PCIe1 and the PCIe switch) 
for RDMA support. 
In comparison, 
SNIC \ding{194}* communication can leverage DMA to reduce such interferences.
For example, for READ with payloads 16--64\,B, 
we only observe a 5--6\% throughput drop,
after adding SNIC \ding{194}* to \ding{192}.

\stitle{Bottleneck.}
Assuming each path has only one type of request, e.g., either READ or WRITE. 
For SNIC \ding{192}+\ding{193}, each part has the same bottleneck (the NIC), 
so the bandwidth limit is 400\,Gbps (bi-directional).
For SNIC \ding{192}+\ding{194}, it is bottlenecked by SNIC \ding{194}, 
which is limited on the uni-direction of PCIe (256\,Gbps) since it occupies 
both directions of PCIe1 (see {\fig{fig:offloading-bottlenecks}(b)}).
Nevertheless, if SNIC \ding{192} is used in opposite directions (i.e., READ 
and WRITE), SNIC \ding{192}+\ding{194} can reach a higher limit. 
For example, the aggregated bandwidth can achieve 456\,Gbps (in theory) if we restrict the bandwidth of 
data transfer on SNIC \ding{194} to 56\,Gbps. 
This suggests that selectively offloading small portion of data to SoC may be optimal.
Finally, if possible, it is usually better to combine SNIC \ding{192} or \ding{193} 
with DMA (\ding{192}/\ding{193}+\ding{194}*) despite DMA being 
slower than RDMA (see \textsection{\ref{sec:dataflow-3}}).
This is because DMA has better PCIe utilization (without passing PCIe) and RNIC utilization 
(without using RNIC).

\stitle{Takeaways.} 
Sending requests from clients to the host and the SoC concurrently 
(SNIC \ding{192}+\ding{193}) can better utilize NIC cores to handle small RDMA requests, 
especially when used in opposition directions (e.g., one for READ and one for WRITE).
On the contrary, uncontrolled use of intra-machine (host-SoC) communications 
(SNIC \ding{194}) may harm 
inter-machine communications, which is the intrinsic purpose of using {\snic}.
Specifically, if the uni-directional bandwidth of PCIe is smaller than 
the bi-directional bandwidth of the NIC, 
using SNIC \ding{194} can introduce a hidden bottleneck.
Therefore, we should always consider using SNIC \ding{194} only when spare resources 
are made available. 
Specifically, if the inter-machine communication saturates the NIC,
the bandwidth used by SNIC \ding{194} should no larger than $P-N$, 
where $P$ and $N$ are the limit of the PCIe and the network, respectively. 
For example, it should be 56\,Gbps on our testbed.
Using SNIC \ding{194}* can reduce the interference between paths, 
but SNIC \ding{194}* also has limitations: it is slower than SNIC \ding{194}.

Finally, in real-world distributed systems, it is common that a single communication path 
cannot fully saturate all resources of {\snic}. For example, SNIC \ding{193} is the fastest 
but limited by small memory and wimpy cores on the SoC. On the other hand, 
only using SNIC \ding{192} as RNIC would waste all resources on the SoC.
Therefore, we should concurrently use multiple paths provided by the {\snic}, 
but carefully avoid interference between them.

\begin{figure*}[!t]
    \hspace{-2mm}
    \begin{minipage}{1\textwidth}
    \centering 
    \includegraphics[scale=1.1]{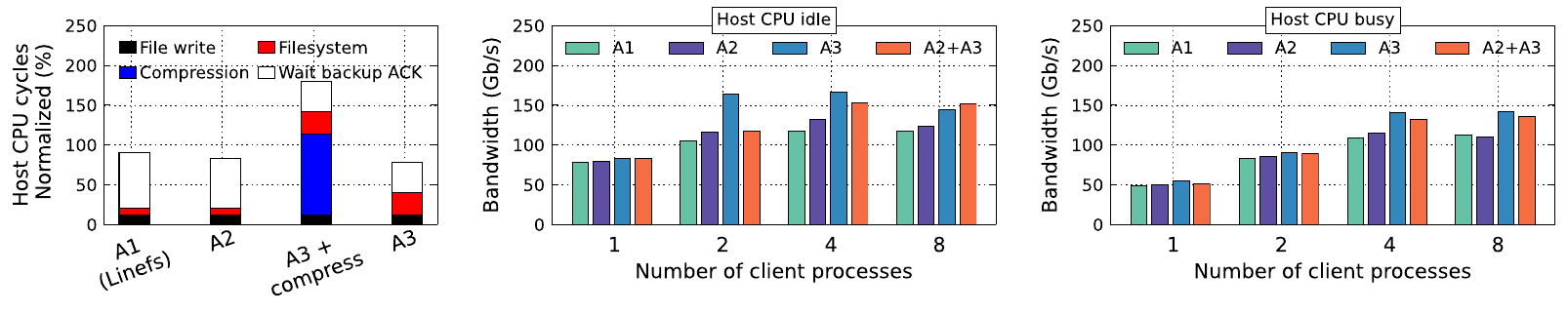}  
    \end{minipage} \\[3pt]
    \begin{minipage}{1\textwidth}
    \caption{\small{%
    The host CPU usage breakdown of different alternatives when replicate 8\,MB data (a). 
    Write bandwidth when the host is idle (b) and busy (c) for A1, A2 and A2 + A3.
    }}
    \label{fig:linefs-data}
    \end{minipage} \\[-20pt]
    \end{figure*} 

\subsection{An optimization guideline}
\label{sec:guideline}

\noindent
This section presents our optimization guideline
for smartly utilizing multiple communication 
paths of {\snic} to improve the performance of distributed systems. 
Specifically, given the functionality (e.g., file replication 
in a distributed file system)  of a target distributed system that needs
to be accelerated by {\snic}, 
we recommend designers consider the following steps: \\[-16pt]
\begin{enumerate}[leftmargin=*,leftmargin=13pt,itemindent=0pt]  
    \item Devise potential alternatives for {\snic} to support the given functionality, 
	and optimize them based on performance characteristics uncovered by our study.     
    \\[-16pt]
    \item Evaluate and rank alternatives based on system-specific criteria. 
    \\[-16pt]
    \item Select and combine alternatives in turn until the resource of {\snic} is saturated. 
    \\[-16pt]
\end{enumerate}

\stitle{System-specific criteria.}
The criteria can be the desirable properties that the system designer aims to achieve,
or the restrictions of the systems. 
For replication in a distributed file system, the properties include low host CPU overhead 
and high network bandwidth utilization~\cite{linefs}. 
For a disaggregated key-value store, the properties include less network amplification, 
low latency and high throughput. 
The restriction the host has little or no CPU that we can use~\cite{DBLP:conf/usenix/ZuoSYZ021}.

\stitle{Discussion.}
We currently only consider the combination of alternatives in a greedy way,
which is sufficient for most networked functions in real-world distributed systems.
Further, {\snic} usually offers a limited number of available options. 
Note that efficiently combining alternatives is challenging. 
For different systems, different alternatives may consume different resources on the {\snic}, 
while a combination of them may involve different levels of resource contentions. 
Our previous analysis---including the bottleneck of different communication paths and concurrently 
utilizing multiple paths on the {\snic}---will guide designers to avoid most performance contention.
Nevertheless,
how to systematically choose and combine different paths is our future work.

\section{Case Studies}
\label{sec:sys}

\noindent
To demonstrate the efficacy of our study and the optimization guideline, 
this section presents two detailed case studies.

\subsection{Distributed file system}
\label{sec:linefs}

\vspace{-2mm}
\stitle{Overview.}
File replication is a key pillar in distributed file systems for fault tolerance.
With the emergence of RDMA and non-volatile memory (NVM), 
an appealing trend is to use RDMA to directly replicate file updates on remote NVM 
for better performance~\cite{linefs, DBLP:conf/osdi/AndersonCKKKPRS20, DBLP:conf/usenix/LuSCL17, Assise},
i.e., RDMA primitives can directly write NVM just like DRAM,
with network and NVM bandwidth fully utilized~\cite{DBLP:conf/usenix/WeiX00Z21}.

\begin{figure}[!t]
\vspace{-2mm}
\begin{minipage}{1\linewidth}
\includegraphics[scale=.95]{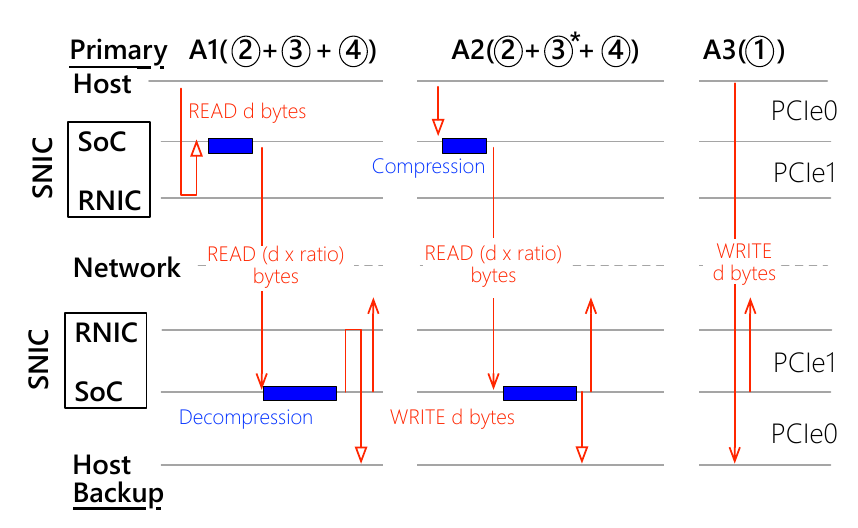}  
\end{minipage} \\[5pt]
\begin{minipage}{1\linewidth}
\caption{\small{%
Overview of alternatives for file replication with {\snic}.
$ratio$ is defined as \texttt{compressed size} / \texttt{uncompressed size}. 
We omit the control path messages as they are trivial. 
}}
\label{fig:linefs}
\end{minipage} \\[-15pt]
\end{figure}

\stitle{Devise alternatives.}
The desirable properties of file replication are high performance, 
high network utilization and low host CPU overhead. 
There are three alternatives to implement file replications on our {\snic}, 
as illustrated in {\fig{fig:linefs}}. \\[-16pt]

\begin{enumerate}[leftmargin=*,leftmargin=13pt,itemindent=0pt]  
    \item \textbf{Alternative (A1).} 
    It comes from the state-of-the-art distributed file system on {\snic}, 
    {\linefs}~\cite{linefs},
    which completely offloads the file replication to SoC.
    The SoC will compress and replicate the file to reduce data transferred 
	through the network with low host CPU usage. After receiving a replication request, 
    the primary SoC reads the file from host (\ding{194}), compresses it (\ding{195}), 
	and writes the file to remote backups with chain replication~\cite{DBLP:conf/osdi/RenesseS04} (\ding{193}). 
    Specifically, 
    if there are multiple backups, the second backup will further re-replicate the log to the next backup on the chain and so on. \\[-16pt]

    \item \textbf{Alternative (A2).} 
    Guided by our study, 
    we can replace the \ding{194} in A1 with \ding{194}* 
    to reduce interference on the PCIe bandwidth, 
    specifically, PCIe1 on the {\snic}. \\[-16pt]
    
    \item \textbf{Alternative (A3).} The host can directly write the file from the host 
	to the remote backup with WRITE (\ding{192})~\cite{DBLP:conf/usenix/LuSCL17}.
    Note that this approach typically skips file compression 
    to prevent non-trivial host CPU overhead (see {\fig{fig:linefs-data} (a)}). \\[-16pt]
\end{enumerate}

\stitle{Baseline.} 
{\linefs}~\cite{linefs} is a state-of-the-art distributed file system 
based on NVM and {\snic}.
It adopts A1 to replicate the files.
We further implement A2 and A3 on its open-source codebase\footnote{\footnotesize{\burl{https://github.com/casys-kaist/LineFS}}},
and rewrite its backend with more efficient RDMA implementation to scale to 200\,Gbps networking,
e.g., with asynchronous and batched RDMA operations.

\stitle{Optimization on each alternative.}
By default, {\linefs} adopts a chunk size of 16\,MB in its open-source codebase for A1.
Based on our Advice\,\#3 described in \textsection{\ref{sec:dataflow-3}},
we shrink it to 256\,KB for a better performance over \ding{194}.
This optimization further applies to A2 and A3. 

\stitle{Analyse alternatives.}
A1 is the most straightforward way to offload file replication, 
reducing the data transferred through the network ($d$ vs. $d \times ratio$). 
Thus, the ideal peak bandwidth is $N / ratio$, where $N$ is the bandwidth limit of {\snic}.
However, 
A1 does not consider the costly PCIe occupation of \ding{194} (\textsection{\ref{sec:dataflow-3}}), 
which even fails to saturate the network bandwidth for file transfer.
Denote the primary's PCIe limit (uni) as $P$. 
A1's file transfer bandwidth $d$ is limited by 
$\frac{P}{1 + ratio}$, 
because each data packet must pass the \emph{PCIe1 out link} twice. 
As shown in {\fig{fig:linefs}}, 
one is from SoC to RNIC ($d$ bytes) and another from SoC to the remote ($d \times ratio$ bytes). 
On our platform ($p=256\,Gbps$),
so A1 is only better than file is not compressed (whose performance is bottlenecked by the network $N=200\,Gbps$) 
when the compression ratio is lower than 28\%.
Worse even, A1 cannot saturate the network bandwidth of {\snic} 
when encountering a bad compression ratio ($\geq$ 28\%). 
For example, without compression ($ratio=1$), the peak of A1 is only 128\,Gbps.

{\fig{fig:linefs-data}} (b) presents the results of A1 on the file write benchmark of {\linefs}.
This benchmark does not compress the file.
We can see that A1 only achieves 
117\,Gbps with 8 clients when the host is idle.

A2 addresses the poor PCIe utilization of A1 by replacing \ding{194} with \ding{194}*.
As shown in {\fig{fig:linefs-data}} (b), A2 is 1.01--1.13\,$\times$ faster than A1 
under different number of clients.
However, A2 fails to achieve a close to 200\,Gbps result (peak at 133\,Gbps) due to the following two reasons. 
First, the WRITE of \ding{194}* cannot fully utilize the full PCIe bandwidth on our platform (see {\fig{fig:rdma-vs-dma}}).
Second, the poor computation power of SoC may also become the performance bottleneck of file replication.

A3 bypasses the PCIe occupation problem of A1, 
and the slow DMA WRITE and weak SoC issues of A2.
Meanwhile, its data path is shorter (see {\fig{fig:linefs}}).
As shown in {\fig{fig:linefs-data}}(a), it takes 40\% shorter time 
to wait for the log acknowledgment compared to A2.
As a result, A3's replication bandwidth is 5--41\% faster than A2 under different client setups.
The drawback is that A3 takes more CPU cycles even without considering compression (Filesystem), 
see {\fig{fig:linefs-data}}(a). 
This is because A1 and A2 can digest the file log on the SoC. 
Thus, the overall process time reduction of A3 is 8\% (decreased from 40\%) compared to A2.

\begin{figure}[!t]
\begin{minipage}{1\linewidth}
\centering 
\includegraphics[scale=1.1]{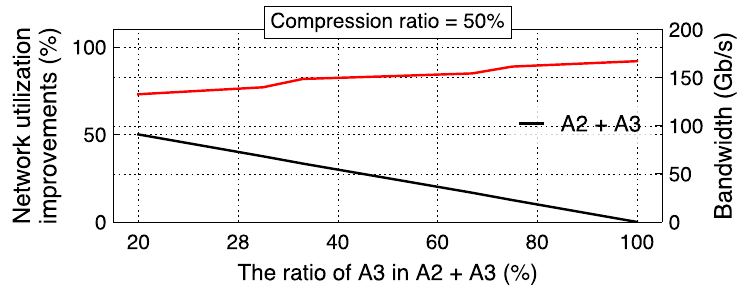}  \\[2pt]
\end{minipage} \\[5pt]
\begin{minipage}{1\linewidth}
\caption{\small{%
Analysis of the network utilization and performance when combining A2 and A3. 
}}
\label{fig:linefs-analysis}
\end{minipage} \\[-15pt]
\end{figure}

\stitle{Select and combine alternatives.}
Since A2 is always better than A1, 
we will only consider combining A2 with A3. 
As we have analyzed before, A3 is faster than A2.
Therefore, increasing the ratio of A3 in a combined path (A2 + A3) always improves the performance, 
as shown in {\fig{fig:linefs-analysis}}.
However, if file compression for high network utilization is enabled, 
it has high host CPU utilization, as shown in {\fig{fig:linefs-data}} (a). 
Disabling compression for A3 will lower the network utilization, 
also illustrated in {\fig{fig:linefs-analysis}}.
Specifically, when increasing the percentage of path A3 in clients, 
the network utilization is reduced from 50\% to 0\% considering a fixed 50\% compression ratio.

Considering A2 has better network utilization, 
we follow a greedy approach that first saturate the SoC with A2 for better network utilization. 
Afterward, 
clients use A3 to do the file replication.
This approach can achieve the best of both worlds: 
the combined path is faster than A2 with network better utilized than A3.

\stitle{Evaluation results.}
{\fig{fig:linefs-data}} (b) and (c) further present the file replication benchmark results 
of A2 + A3 when the host CPU is idle and busy, respectively.
We follow the same setup as {\linefs}~\cite{linefs}'s benchmark and 
add a CPU-intensive workload (streamcluster~\cite{DBLP:conf/IEEEpact/BieniaKSL08}) to the host CPU
to emulate a busy experimental setup. 
A2 + A3 is 7--30\% and 4--21\% faster than original {\linefs}
when CPU is idle and busy, respectively,
thanks to the more efficient usage of {\snic} and a smart utilization of multiple execution paths.

\begin{figure*}[!t]
\hspace{-3mm}
\centering 
\includegraphics[scale=0.77]{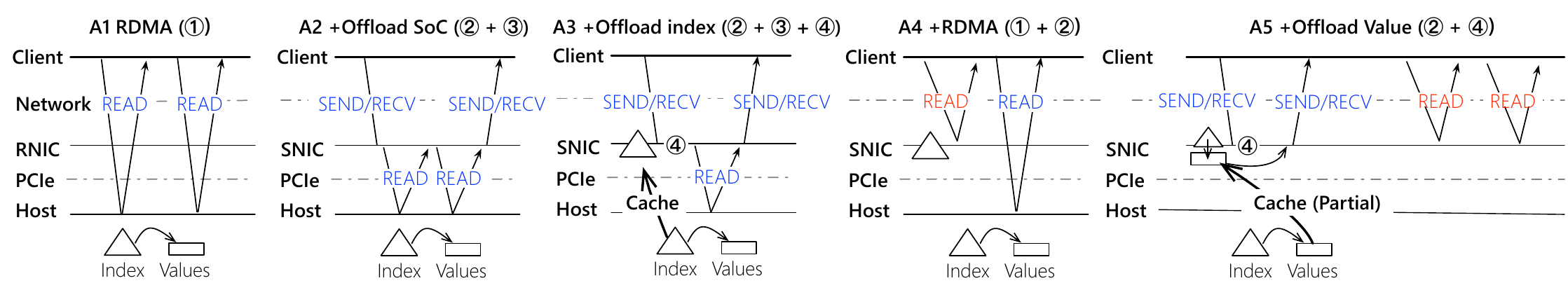}  \\[0pt]
\begin{minipage}{1\linewidth}
\caption{\small{%
Alternatives (A1--A5) for offloading a get request of RDMA-based 
disaggregated key-value store to off-path {\snic}. 
}}
\label{fig:kv-offload}
\end{minipage} \\[-0pt]
\end{figure*}

\subsection{Disaggregated key-value store}
\label{sec:kvs}

\vspace{-2mm}
\stitle{Overview.}
RDMA-based disaggregated key-value stores (\rkv) are prevalent 
in modern data centers~\cite{xstore,DBLP:conf/usenix/TsaiSZ20,farm,DBLP:conf/usenix/ZuoSYZ021}. 
In {\rkv}, 
one or more memory servers store both indexes (usually hash table) and values. 
Clients on other machines use READs to traverse the index and retrieve 
the corresponding value to handle requests (i.e., get), see A1 in {\fig{fig:kv-offload}}. 

\stitle{Devise alternatives.}
The desired properties 
are high throughput, low latency and minimal network amplification. 
The restriction is that we can barely use the host CPU (i.e., disable SEND/RECV for path \ding{192}). 
{\snic} enables five alternatives for {\rkv},
as illustrated in {\fig{fig:kv-offload}.  \\[-16pt]

\begin{enumerate}[leftmargin=*,leftmargin=13pt,itemindent=0pt]  
    \item \textbf{Alternative (A1).} 
	The client treats {\snic} as a normal RNIC and uses READs to handle the {get} request (\ding{192}). 
    This approach suffers from network amplification. \\[-16pt] 

    \item \textbf{Alternative (A2).} 
    One intuitive approach for offloading is to send the {get} request to the SoC 
	using SEND/RECV (\ding{193}).     
    The SoC can then traverse the index and read the value on the host via RDMA or DMA READ. 
    This approach effectively eliminates network amplification.  \\[-16pt]

    \item \textbf{Alternative (A3).} 
    One drawback of A2 is that reading data from SoC to the host is slower reading from the host's local memory. 
    An optimization is to offload the indexes to the SoC memory (\ding{195}).
    This approach is similar to index caching at the clients~\cite{drtm-tocs, farm-v2, xstore},
    but caching the indexes at the {\snic} is more effective. 
    Each client has a small memory that can only cache hundreds of entries 
	in a disaggregated setting~\cite{DBLP:conf/usenix/ZuoSYZ021}, 
    while {\snic} has a relatively large SoC memory (e.g., 16\,GB on {\dpu}-2) that can
	cache all the indexes. \\[-16pt]

    \item \textbf{Alternative (A4).}
    Accessing the index on the SoC using SEND/RECV (\ding{193}) cannot fully utilize 
	the NIC cores of {\snic}, because the peak throughput of SEND/RECV is only 21.6\,M reqs/s. 
    Therefore, we can use READs to traverse the index on the SoC (\ding{193}), 
    and another READ to retrieve the value on the host (\ding{192}). 
    This approach still has network amplification, but can utilize the fast path 
	(\ding{193}) to improve performance (see \textsection{\ref{sec:dataflow-2}}).  \\[-16pt]

    \item \textbf{Alternative (A5).} 
    Similar to index caching, SoC memory can further cache a portion of values (e.g., 
	the values of hot keys). This approach avoids using the costly communication path 
	(\ding{194}) of the previous alternatives. \\[-16pt]     
\end{enumerate} 

\begin{figure}[!t]
\begin{minipage}{1\linewidth}
\hspace{-5pt}
\includegraphics[scale=1.2]{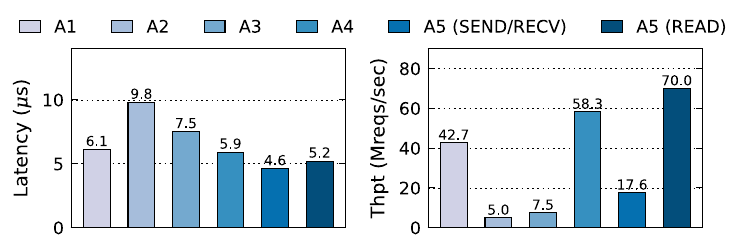}
\end{minipage} \\[5pt]
\begin{minipage}{1\linewidth}
\caption{\small{%
(a) Latency and (b) throughput comparisons between different alternatives on YCSB C. 
For A5, we restricted the selection of clients keys to always hit the cached values on {\snic}. 
}}
\label{fig:kv-data}
\end{minipage} \\[-0pt]
\end{figure} 

\stitle{Baseline.} 
{\drtm}~\cite{drtm-tocs} is a state-of-the-art KV store optimized for RDMA: 
it adopts cluster-chaining hash index such that the client typically finds 
the value position of a given key in one READ.
Specifically, for a {get} request, the client first READs a 64\,B bucket 
(based on the hash of the key), 
finds the remote address of the corresponding value in it, and then fetches the value 
with another READ. 
{\drtm} supports index caching at the client to skip the first READ~\cite{DBLP:conf/sosp/WeiSCCC15}, 
but it may not be always feasible in a disaggregated environment due to memory constraints~\cite{DBLP:conf/usenix/ZuoSYZ021}, 
so we disable it. 

\stitle{Optimization on each alternative.}
We implement A1--A5 on {\drtm} guided by our study (\textsection{\ref{sec:dpu-study}}). 
Specifically, we carefully enabled doorbell batching for alternatives related to the SoC CPU (A2, A3 and A5). 
Besides, we apply Advice\,\#1 for A4 and A5, 
which replicate a few hot keys to multiple replications to tame the skewness. 
We use DMA (\ding{194}$^*$) instead of RDMA (\ding{194}) to implement A2 and A3 as it is always faster due to lower latency. 
For example, A2 throughput is improved  by up to 79\% with \ding{194}*.
A2 and A3 does not suffer from the low DMA throughput discovered in \textsection{\ref{sec:rdma-vs-dma}} because the SoC 
will first become the bottleneck.

\stitle{Analyse alternatives.}
We use YCSB C~\cite{ycsbgit} (100\% {get}) with default Zipfian request distribution 
($\theta = 0.99$) for all the experiments. 
The payload sizes of keys and values are 8\,B and 64\,B, respectively, 
similar to prior work~\cite{DBLP:conf/eurosys/MaoKM12,DBLP:conf/usenix/MitchellMNSL16,erpc,tang2020xindex,xstore}.
Following the microbenchmark setup, 
we use one client machine to measure the latency 
and deploy up to eleven client machines to measure the peak throughput. 

{\fig{fig:kv-data}} demonstrates that none of the path can achieve 
both high throughput and low latency. 
A5 (SEND/RECV) achieves the lowest latency (4.6\,$\mu$s) 
because it completely eliminates the network amplifications problem 
and costly host-SoC communications (\ding{194}). 
However, its peak throughput (17.6\,M reqs/s) is significantly lower
than some other alternatives.
Specifically, the peak throughput of A5 (READ) and A4 reach 70\,M reqs/s 
and 58.3\,M reqs/s, respectively. 
They have a higher throughput because the RDMA path to SoC (\ding{192}) 
is faster (\textsection{\ref{sec:dataflow-2}}). 
Note that A5 is not always achievable, which requires caching all the key-values at the SoC memory.
Therefore, A4 is a suitable design if the SoC cores become the bottleneck (\ding{195}).
A1 has a higher latency and lower throughput than A4, since RDMA to 
the host (\ding{192}) is relatively slow. 
A2 and A3 are bottlenecked by the slow host-SoC communication (\ding{194}, 
see \textsection{\ref{sec:dataflow-3}}), 
which is not suitable for offloading KV store requests.

\stitle{Select and combine alternatives.}
Our analysis suggests that the optimal combination is A4 and A5. 
Initially, 
the first few clients use A5, whereas the later clients use A4. 
The exact switch point can be estimated by using queuing theory~\cite{harchol2013performance} 
to model the capacity of SoC and the capability of RNIC, as in prior work~\cite{cell}. 

In addition, using A5 presents a challenge as clients are unaware of which values 
are cached at SoC. Although A3 can be used as a fallback path for cache misses, 
it will result in significant performance degradation (see {\fig{fig:kv-data}}). 
To tackle this issue, we provide a simple solution:
when a cache miss occurs, the SoC returns the address of the value to the client, 
which then issues a READ to retrieve the value accordingly, similar to A4.
In real-world skewed workloads (e.g., YCSB~\cite{ycsbgit}), cache misses are rare. 

\begin{figure}[!t]
\vspace{-1mm}
\hspace{-3mm}
\centering 
\includegraphics[scale=1.2]{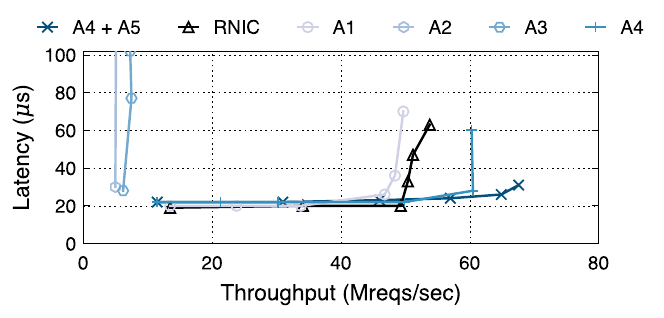}  \\[7pt]
\begin{minipage}{1\linewidth}
\caption{\small{%
Performance of YCSB C using different alternatives.  
Note that A5 cannot run a full workload alone, since SoC memory is not large 
enough to cache all values.
For A4 + A5, one client uses A5, and the rest use A4. 
}}
\label{fig:kvs-thpt-lat}
\end{minipage} \\[-15pt]
\end{figure}

\stitle{Evaluation results.}
{\fig{fig:kvs-thpt-lat}} shows the latency and throughput results on YCSB C.
We plotted the graph by increasing the number of client machines. 
The combination of A4\,+\,A5 achieves a peak throughput of 68\,M reqs/s,
which is 25\%, 36\%, and 12\% higher than RNIC, A1, and A4, respectively. 
Note that we omit A2 and A3 as they 
are bottlenecked by SoC cores and have extremely low peak throughput. 
The benefits of A4\,+\,A5 mainly come from utilizing faster SoC RDMA and SoC cores 
for reducing network amplifications.

\section{Discussion}
\label{sec:dis}

\vspace{-2mm}
\stitle{Generalizability.} 
Although our study primarily focuses on one particular {\snic}, {\dpu}-2~\cite{bluefield-2}, 
we believe that our findings and advice can be applied to other off-path {\snic}s 
that share a similar hardware architecture. These {\snic}s extend RDMA-capable NICs, 
such as Stingray PS225~\cite{broadcom-stingray} 
(which extends NetXtreme 100\,Gbps RNIC~\cite{bcm57504-100g-ic}), 
by attaching a heterogeneous SoC and bridging SoC and RNIC together with a PCIe switch. 
We have confirmed that all our results hold on {\dpu}-1~\cite{bluefield-1}.
Moreover, the next generation of {\dpu} ({\dpu}-3) still follows the same architecture, 
except for using faster RNIC (400\,Gbps ConnectX-7), PCIe (5.0), and SoC (ARMv8.2+ A78).
Even though other {\snic}s may have different parameters than {\dpu}-2, 
our methodology, analysis tools (open-sourced), and 
performance models (e.g., Table~\ref{tab:pcie-payload}) also apply to them.

Furthermore, DPDK~\cite{dpdk} is another popular communication primitive over {\snic}. 
From a NIC's perspective, DPDK is similar to SEND/RECV over UD. 
Therefore, we believe that most of our findings are still applicable to DPDK as well. 
Unfortunately, we do not have an Ethernet-based testbed to confirm this further. 

\stitle{Suggestions for hardware vendors.}
Our study has uncovered several anomalies that can be mitigated through hardware improvements, 
which we suggest vendors consider. 
For example,
current host to SoC DMA must offload to SoC for execution~\cite{doca-dma-programming},
while supporting CXL~\cite{cxl} can utilize the more powerful host CPU DMA engine for it.
However, doing so in a programmer-friendly way~\cite{280792} will require 
strong cooperation between the SoC OS and host OS. 
To the best of our knowledge, no {\snic} supports CXL yet. 
Moreover, vendors can support CCI~\cite{cci} to mitigate the write skew problem 
mentioned in \textsection{\ref{sec:dataflow-2}}.  
Moreover, aligning the SoC PCIe MTU with the host is likely to improve PCIe performance 
when transferring large payloads. 
Finally, we encourage vendors to disclose more hardware details of {\snic}s
to help explain and confirm the findings of our study.

\section{Other Related Work}
\label{sec:related}

\vspace{-2mm}
\stitle{SmartNIC offloading.}
Offloading computation to {\snic}s has attracted significant attention in academia and industry. 
The offloaded tasks include network functions~\cite{floem,clicknp,DBLP:conf/nsdi/FirestonePMCDAA18}, 
microservices~\cite{lambda-nic, e3}, and others~\cite{kv-direct, panic,fairnic,ainico,xenic,DBLP:conf/usenix/WangH00A22,DBLP:conf/eurosys/SidlerWCKA20}. 
We share the same vision---improving the performance of distributed systems 
by offloading computation and communication to {\snic}s, but further exploit 
the multiple communication paths of {\snic}s.
In addition, most prior work has focused on leveraging a single path of on-path {\snic}s, 
so our work 
can inspire future research on multi-path offloading for on-path {\snic}s.

\stitle{RDMA offloading.}
Before the emergence of {\snic}s, many distributed systems offloaded remote memory accesses 
to one-sided RDMA primitives~\cite{xstore,wukong,farm,farmv2,cell,storm-rdma-tx,drtmr,drtm-h,
xie2019pragh,wei2017drtmb,DBLP:journals/pvldb/ZamanianBKH17,DBLP:conf/usenix/LuSCL17,
zhang2017wukongs,DBLP:conf/usenix/ZuoSYZ021}. 
However, prior work has observed the poor semantics of one-sided RDMA and 
has therefore leveraged advanced RDMA features 
(e.g., WAIT~\cite{DBLP:journals/corr/abs-2103-13351, DBLP:conf/sigcomm/KimMBZLPRSSS18},
DCT~\cite{krcore,mitosis}) 
or introduced new RDMA primitives~\cite{DBLP:conf/eurosys/SidlerWCKA20,DBLP:conf/sosp/0001DJSNZP21}. 
These efforts are orthogonal to our work and could also benefit from our findings 
when using {\snic}s in the future.

\section{Conclusion}

\noindent
Designing high-performance distributed systems with {\snic}s requires 
an in-depth understanding of low-level hardware details. 
This paper presents a comprehensive study of off-path {\snic}. 
Unlike prior work, we explore 
how the {\snic} architecture and the heterogeneity of its computation units 
can impact communication performance related to its components. 
We further propose the first optimization guideline for designers to smartly 
exploit multiple communication paths of {\snic}s for distributed systems,
and demonstrate our guideline by improving two distributed systems. 
In general, 
our study can help system designers develop a better understanding of {\snic}s 
before applying them in high-performance distributed systems. 

\section*{Acknowledgment}

\noindent
We sincerely thank our shepherd, Rachit Agarwal, and the anonymous reviewers
for their comments and suggestions to improve the paper. 
We also thank 
Zhuobin Huang for discussing DPDK on {\snic}, 
Jun Lu and Dong Du for providing the Bluefield-1 hardware, 
Dingji Li, Erhu Feng, Jinyu Gu, and Fangming Lu for their valuable feedback 
on earlier versions of the paper. 
This work was supported in part by 
the National Key Research \& Development Program of China (No. 2022YFB4500700), 
the Fundamental Research Funds for the Central Universities,
the National Natural Science Foundation of China (No. 62202291, 62272291, 61925206), 
as well as research grants from Huawei Technologies and Shanghai AI Laboratory. 
Corresponding author: Rong Chen (\burl{rongchen@sjtu.edu.cn}).

\balance

\small{
\bibliographystyle{acm}
\bibliography{sx}
}

\clearpage

\end{document}